\documentclass[11pt,a4paper]{article}
\usepackage{amsmath,latexsym,fullpage,amssymb,color}
\usepackage{epic,eepic,ifthen,graphics,epsfig}
\usepackage[english]{babel}
\usepackage{times}
\usepackage{float}
\usepackage{xspace}
\usepackage[T1]{fontenc}
\usepackage{tikz}

\newlength {\squarewidth}

\newtheorem{definition}{Definition}
\newtheorem{theorem}{Theorem}
\newtheorem{lemma}{Lemma}
\newtheorem{corollary}{Corollary}

\newcommand{\toto}{xxx}
\newenvironment{proofT}{\noindent{\bf Proof }}
{\hspace*{\fill}$\Box_{Theorem~\ref{\toto}}$\par\vspace{3mm}}
\newenvironment{proofL}{\noindent{\bf Proof }}
{\hspace*{\fill}$\Box_{Lemma~\ref{\toto}}$\par\vspace{3mm}}

\newenvironment{lemma-repeat}[1]{\begin{trivlist}
\item[\hspace{\labelsep}{\bf\noindent Lemma~\ref{#1} }]}%
{\end{trivlist}}

\newcounter{linecounter}
\newcommand{\linenumbering}{\ifthenelse{\value{linecounter}<10}
{(\arabic{linecounter})}{(\arabic{linecounter})}}
\renewcommand{\line}[1]{\refstepcounter{linecounter}\label{#1}\linenumbering}
\newcommand{\resetline}[1]{\setcounter{linecounter}{0}#1}
\renewcommand{\thelinecounter}{\ifnum \value{linecounter} > 
9 \else \fi\arabic{linecounter}}
\newfloat{algorithm}{thp}{lop}
\floatname{algorithm}{Algorithm}
\newfloat{construction}{thp}{lop}
\floatname{construction}{Construction}
\newfloat{pattern}{thp}{lop}
\floatname{pattern}{Pattern}

\newcommand{\Xomit}[1]{}

\newcommand{\REG}{\mathit{REG}}

\newcommand{\lub}{{\sf lub}}
\newcommand{\op}{{\sf op}}
\newcommand{\progress}{{\sf progress}}
\newcommand{\forward}{{\sf forward}}
\newcommand{\trydeliver}{{\sf try\_deliver}}
\newcommand{\ttrue}{\tt{true}}
\newcommand{\ffalse}{\tt{false}}
\newcommand{\done}{\mathit{done}}

\newcommand{\SYNC}{{\sc{sync}}}
\newcommand{\MSG}{{\sc{msg}}}
\newcommand{\TYPE}{{\sc{type}}}
\newcommand{\WRITE}{{\sc{write}}}
\newcommand{\FORWARD}{{\sc{forward}}}
\newcommand{\reg}{\mathit{reg}}
\newcommand{\buffer}{\mathit{buffer}}
\newcommand{\bufferi}{\mathit{buffer_i}}    
\newcommand{\bufferj}{\mathit{buffer_j}}

\newcommand{\snapshot}{{\sf snapshot}}
\newcommand{\wwrite}{{\sf write}}
\newcommand{\catchup}{{\sf catchup}}
\newcommand{\rread}{{\sf read}}
\newcommand{\propose}{{\sf propose}}

\newcommand{\increase}{{\sf increase}}
\newcommand{\decrease}{{\sf decrease}}

\newcommand\past{\mathit{past}}

\newcommand{\wait}{{\sf{wait}}}
\newcommand{\return}{{\sf{return}}}

\newcommand{\mmin}{{\sf min}}

\newcommand{\snsd}{{\mathit{sn_{sd}}}}
\newcommand{\snf}{{\mathit{sn_{f}}}}

\newcommand{\SETSEQ}{\mathit{SETS\_SEQ}}
\newcommand{\setseq}{\mathit{sets\_seq}}
\newcommand{\SENT}{\mathit{SENT}}
\newcommand{\sent}{{\mathit{sent}}}
\newcommand{\send}{{\sf send}}
\newcommand{\members}{{\sf{members}}}

\newcommand{\entermutex}{\sf{enter\_mutex}}
\newcommand{\exitmutex}{\sf{exit\_mutex}}
\newcommand{\todeliver}{\mathit{to\_deliver}}
\newcommand{\todeliveri}{\mathit{to\_deliver_i}}

\newcommand{\scddelivered}{\mathit{scd\_delivered}}

\newcommand{\CAMP}{{\cal{CAMP}}}
\newcommand{\CARW}{{\cal{CARW}}}
\newcommand{\SCD}{{{SCD}}}
\newcommand{\PLUS}{{\sc plus}}
\newcommand{\MINUS}{{\sc minus}}

\newcommand{\scdbroadcast}{{\sf scd\_broadcast}}
\newcommand{\scdeliver}{{\sf scd\_deliver}}
\newcommand{\fifobroadcast}{{\sf fifo\_broadcast}}
\newcommand{\fifodeliver}{{\sf fifo\_deliver}}

\newcommand{\tsa}{\mathit{tsa}}
\newcommand{\TSA}{\mathit{TSA}}
\newcommand{\lexts}{{~<_{ts}~}}
\newcommand{\MS}{\mathit{MS}}

\Xomit{
\newlength{\reduceunderfig}
\setlength{\reduceunderfig}{.2cm}
\usepackage[lmargin=2.1cm,rmargin=2cm,tmargin=2cm,bmargin=2cm]{geometry}
\renewcommand{\baselinestretch}{0.96}
\renewcommand{\paragraph}[1]{\vspace{0.2cm}\noindent \textbf{#1}~~}
\let\subsectionOld\subsection
\renewcommand{\subsection}[1]{\vspace{-0.2cm}\subsectionOld{#1}\vspace{-0.2cm}}
\let\subsubsectionOld\subsubsection
\renewcommand{\subsubsection}[1]{\vspace{-0.3cm}\subsubsectionOld{#1}\vspace{-0.15cm}}
} 

\begin{document}

\title{\bf  Set-Constrained Delivery Broadcast:\\  
             Definition,  Abstraction Power, and Computability Limits}

  \author{Damien Imbs$^{\circ}$, 
         Achour Most\'efaoui$^{\dag}$,~
         Matthieu Perrin$^{\diamond}$,~
         Michel Raynal$^{\star,\ddag}$\\~\\
$^{\circ}$LIF, Universit\'e Aix-Marseille, 13288  Marseille, France \\
$^{\dag}$LINA, Universit\'e de Nantes, 44322 Nantes, France \\
        $^{\diamond}$IMDEA Software Institute, 28223 Pozuelo de Alarc\'on,
        Madrid,  Spain\\
$^{\star}$Institut Universitaire de France\\
$^{\ddag}$IRISA, Universit\'e de Rennes, 35042 Rennes, France \\
}


\maketitle


\begin{abstract}
This paper  introduces a new communication abstraction, called
{\it Set-Constrained Delivery Broadcast} (SCD-broadcast), whose aim is to
provide its users with an appropriate abstraction level when they have
to implement {\it objects} or {\it distributed tasks} 
in an asynchronous message-passing system prone to process crash failures.
This abstraction allows each process to broadcast messages and deliver a
sequence of sets of messages in such a way that, if a process delivers a set
of messages including a message $m$ and later delivers a set of messages
including a message $m'$, no process delivers first a  set of messages
including  $m'$ and later a set of message including $m$.

After having presented  an algorithm implementing SCD-broadcast,
the paper investigates its programming power and its computability
limits.  On the ``power'' side it presents SCD-broadcast-based
algorithms, which are both simple and efficient, building objects (such as
snapshot and conflict-free replicated data), and distributed tasks. 
On the ``computability limits'' side it shows that
SCD-broadcast and read/write registers are computationally equivalent.
~
\end{abstract}

\thispagestyle{empty}
\newpage
\setcounter{page}{1}

\section{Introduction}

\paragraph{Programming abstractions}
Informatics is a science of abstractions, and a main difficulty consists in 
providing users with a ``desired level of abstraction and generality
-- one that is broad enough to encompass interesting new situations, yet
specific enough to address the crucial issues'' as expressed in~\cite{FM03}. 
When considering sequential computing, functional programming and
object-oriented programming are well-know examples of what means
``desired level of abstraction and generality''.

In the context of asynchronous distributed systems where the computing
entities (processes) communicate --at the basic level-- by sending and
receiving messages through an underlying communication network, and
where some of them can experience failures, a main issue consists in
finding appropriate communication-oriented abstractions, where the
meaning of the term ``appropriate'' is related to the problems we
intend to solve.  Solving a problem at the send/receive abstraction
level is similar to the writing of a program in a low-level
programming language.  Programmers must be provided with abstractions
that allow them to concentrate on the problem they solve and not on
the specific features of the underlying system.  This is not
new. Since a long time, high level programming languages have proved the
benefit of this approach.  From a synchronization point of view, this approach
is the one promoted in {\it software transactional memory}~\cite{ST97},
whose aims is to allow programmers to focus on the synchronization
needed to solve their problems and not on the way this synchronization
must be implemented (see the textbooks~\cite{HS08,R13-1}).

If we consider specific coordination/cooperation problems, 
``matchings'' between problems and specific communication
abstractions are known.
One of the most famous examples concerns the consensus problem whose
solution rests on the {\it total order broadcast} abstraction\footnote{Total
  order  broadcast is also called {\it atomic broadcast}.
  Actually,  total order broadcast and consensus  have been shown to be
  computationally equivalent~\cite{CT96}. A more general result is presented
  in~\cite{IMPR17}, where is introduced a communication abstraction
  which ``captures'' the $k$-set agreement problem~\cite{C93,R16}
  (consensus is $1$-set agreement).}.  Another ``matching'' example 
is the {\it causal message delivery} broadcast abstraction~\cite{BJ87,RST91},
which allows for a very simple implementation of a causal read/write
memory~\cite{ANBHK95}.

\paragraph{Aim  of the paper}
The aim of this paper is to introduce and investigate a high level
communication abstraction which allows for simple and efficient
implementations of concurrent objects and distributed tasks, in the
context of asynchronous message-passing systems prone to process crash
failures. The concurrent objects in which we are interested are
defined by a sequential specification~\cite{HW90} (e.g., a queue).
Differently, a task extends to the distributed context the notion of a
function~\cite{BMZ87,MW87}.  It is defined by a mapping from a set of input
vectors to a set of output vectors, whose sizes are the number of
processes.  An input vector $I$ defines the input value $I[i]$ of each
process $p_i$, and, similarly, an output vector $O$ defines the output
$O[j]$ of each process $p_j$.  Agreement problems such as consensus
and $k$-set agreement are distributed tasks. What makes difficult the
implementation of a task is the fact that each process knows only its
input, and, due to net effect of asynchrony and process failures, no
process can distinguish if another process is very slow or crashed.
The difficulty is actually an impossibility for
consensus~\cite{FLP85}, even in a system in which at most one process
may crash.

\paragraph{Content of the paper: a  broadcast abstraction}
The \SCD-broadcast communication abstraction proposed in the paper
allows a process to broadcast messages, and to deliver sets of
messages (instead of a single message) in such a way that, if a
process $p_i$ delivers a message set $ms$ containing a message $m$,
and later delivers a message set $ms'$ containing a message $m'$, then
no process $p_j$ can deliver first a set containing $m'$ and later
another set containing $m$.  Let us notice that $p_j$ is not prevented
from delivering $m$ and $m'$ in the same set.  Moreover,
\SCD-broadcast imposes no constraint on the order in which a process
must process the messages it receives in a given message set.

After having introduced SCD-broadcast, the paper presents an
implementation of it in asynchronous systems where a minority of
processes may crash. This assumption is actually a necessary and
sufficient condition to cope with the net effect of asynchrony and
process failures (see below).  Assuming an upper bound $\Delta$ on
message transfer delays, and zero processing time, an invocation of
SCD-broadcast is upper bounded by $2\Delta$ time units, and $O(n^2)$
protocol messages (messages generated by the implementation algorithm).

\paragraph{Content of the paper: implementing objects and tasks}
Then, the paper addresses two fundamental  issues of SCD-broadcast: its
abstraction power and its computability limits.  As far as its  abstraction
power is concerned, i.e., its ability and easiness to implement atomic
(linearizable) or sequentially consistent
concurrent objects~\cite{HW90,L86} and read/write solvable distributed
tasks,  the paper presents, on the  one side, two algorithms implementing
atomic objects (namely a snapshot object~\cite{AADGMS93,A94}, and  a
distributed increasing/decreasing counter), and, on the other side,
an algorithm solving the lattice agreement task~\cite{AHR95,FRRRV12}.

The two concurrent objects (snapshot and counter) have been chosen
because they are encountered in many applications, and are also good
representative of the class of objects identified in~\cite{AH90}.  The
objects of this class are characterized by the fact that each pair
$\op1$ and $\op2$ of their operations either commute (i.e., in any
state, executing $\op1$ before $\op2$ is the same as executing $\op2$
before $\op1$, as it is the case for a counter), or any of $\op1$ and
$\op2$ can overwrite the other one (e.g., executing $\op1$ before
$\op2$ is the same as executing $\op2$ alone).  Our implementation of
a counter can be adapted for all objects with commutative operations,
and our implementation of the snapshot object illustrates how
overwriting operations can be obtained directly from the SCD-broadcast
abstraction.
Concerning these objects,  it is also shown that a slight change in the
algorithms allows us to obtain implementations (with a smaller cost) in which
the consistency condition is weakened from linearizability to sequential
consistency~\cite{L79}.

In the case of read/write solvable  tasks,  \SCD-broadcast
shows how the concurrency  inherent (but hidden) in a task definition
can be easily mastered and solved.

\paragraph{A distributed software engineering dimension}
All the algorithms presented in the paper are based on the same
communication pattern. As far as objects are concerned, 
the way this communication pattern is used brings to light
two genericity dimensions of the algorithms implementing them.
One is on the variety of objects that, despite their
individual features (e.g., snapshot vs counter), have very similar
SCD-broadcast-based implementations (actually, they all have the same
communication pattern-based structure). The other one
is on the consistency condition they have to satisfy (linearizability
vs sequential consistency).

\paragraph{Content of the paper: the computability limits of SCD-broadcast}
The paper also investigates the computability power of the
SCD-broadcast abstraction, namely it shows that SCD-broadcast and
atomic read/write registers (or equivalently snapshot objects) have
the same computability power in asynchronous systems prone to process
crash failures.  Everything that can be implemented with atomic
read/write registers can be implemented with SCD-broadcast, and vice versa.

As read/write registers (or snapshot objects) can be implemented in
asynchronous message-passing system where only a minority of processes
may crash~\cite{ABD95}, it follows that the proposed algorithm
implementing SCD-broadcast is resilience-optimal in these systems.
From a theoretical point of view, this means that the consensus number
of SCD-broadcast is $1$ (the weakest possible).

\paragraph{Roadmap}
The paper is composed of~\ref{sec:conclusion} sections.
Section~\ref{sec:scd-broadcast-definition} defines the
SCD-broadcast abstraction and the associated communication pattern
used in all the algorithms presented in the paper.
Section~\ref{sec:SCD-implementation} presents a resilience-optimal
algorithm implementing SCD-broadcast in asynchronous message-passing
systems prone to process crash failures, while
Section~\ref{sec:communication-pattern} adopts a distributed
software engineering point of view and presents a communication pattern
associated with SCD-broadcast. 
Then, Sections~\ref{sec:SCD-power-snapshot}-\ref{sec:SCD-power-LA}
present SCD-broadcast-based algorithms for concurrent objects and tasks. 
Section~\ref{sec:SCD-computability} focuses on the computability limits of
SCD-broadcast. Finally, Section~\ref{sec:conclusion} concludes the paper.

\section{The SCD-broadcast Communication Abstraction}
\label{sec:scd-broadcast-definition}

\paragraph{Process model}
The computing model is composed of a set of $n$ asynchronous
sequential processes, denoted $p_1$, ..., $p_n$. ``Asynchronous'' means
that each process proceeds at its own speed, which can be arbitrary
and always remains  unknown to the other processes.  

A process may halt prematurely (crash failure), but it executes its
local algorithm correctly until it crashes (if it ever does). The model
parameter $t$ denotes the maximal number of processes that may crash
in a run $r$.  A process that crashes in a run is said to be {\it faulty}
in $r$. Otherwise, it is {\it non-faulty}.

\paragraph{Definition of  SCD-broadcast}

The set-constrained broadcast abstraction (\SCD-broadcast) provides
the processes with two operations, denoted $\scdbroadcast()$ and
$\scdeliver()$.  The first operation takes a message to broadcast as
input parameter.  The second one returns a non-empty set of messages
to the process that invoked it.  Using a classical terminology, when a
process invokes $\scdbroadcast(m)$, we say that it ``scd-broadcasts a
message $m$''. Similarly, when it invokes $\scdeliver()$ and obtains a
set of messages $ms$, we say that it ``scd-delivers the set of
messages $ms$''. By a slight abuse of language, when we are interested
in a message $m$, we say that a process ``scd-delivers the message
$m$'' when actually it scd-delivers the message set $ms$ containing $m$.

SCD-broadcast is defined by the following
set of properties, where we assume --without loss of generality--
that all the messages that are scd-broadcast are different. 
\begin{itemize}
\vspace{-0.1cm}
\item Validity.
  If a process scd-delivers a set containing a message $m$,
  then $m$ was scd-broadcast by some process.
\vspace{-0.2cm}
\item Integrity.
  A message is scd-delivered at most once by each process.
\vspace{-0.2cm}
\item MS-Ordering.
  Let $p_i$ be a process that scd-delivers first a message set $ms_i$ and later 
  a message set $ms_i'$. For any pair of messages $m\in ms_i$ and $m'\in ms_i'$,
  then no process $p_j$  scd-delivers first a message set $ms'_j$
  containing $m'$ and later a message set $ms_j$ containing $m$. 
\vspace{-0.2cm}
\item Termination-1.
  If a non-faulty process scd-broadcasts a message $m$,
  it terminates its scd-broadcast invocation and 
  scd-delivers a message set containing $m$.
\vspace{-0.2cm}
\item Termination-2.
  If a process scd-delivers a message $m$, every  non-faulty process
  scd-delivers a message set containing $m$.
\end{itemize}

Termination-1 and Termination-2 are classical liveness properties
(found for example in Uniform Reliable Broadcast~\cite{AW04,R10}). 
The other ones are safety properties.
Validity and Integrity are classical communication-related properties.
The first  states that there is neither  message creation nor message
corruption, while the second states that there is no message duplication.

The MS-Ordering property is new,
and characterizes \SCD-broadcast. It states that the contents of the sets of
messages  scd-delivered at any two processes are not totally independent:
the sequence of sets scd-delivered at a process $p_i$ and
the sequence of sets scd-delivered at a process $p_j$ must be mutually
consistent in the sense that a process $p_i$ cannot scd-deliver
first $m\in ms_i$ and later $m'\in ms_i'\neq ms_i$, while another process 
$p_j$  scd-delivers first $m'\in ms_j'$ and later $m\in ms_j\neq ms_j'$. 
Let us nevertheless observe that if $p_i$  scd-delivers
first $m\in ms_i$ and later $m'\in ms_i'$,
$p_j$ may scd-deliver $m$ and $m'$ in the same set of messages. 

Let us remark that, if the MS-Ordering property is suppressed and
messages are scd-delivered one at a time,  \SCD-broadcast boils down to the
well-known {\it Uniform Reliable Broadcast} abstraction~\cite{CT96,R10}.

\paragraph{An example}
Let $m_1$, $m_2$, $m_3$,  $m_4$, $m_5$, $m_6$, $m_7$, $m_8$, ... be
messages that have been scd-broadcast by different processes.
The following  scd-deliveries of message sets by $p_1$, $p_2$ and $p_3$
respect the definition of \SCD-broadcast: 
\begin{itemize}
\vspace{-0.2cm}
\item at $p_1$:
$\{m_1,m_2\}$, $\{m_3,m_4,m_5\}$, $\{m_6\}$, $\{m_7,m_8\}$.
\vspace{-0.2cm}
\item at $p_2$:
$\{m_1\}$, $\{m_3,m_2\}$, $\{m_6,m_4,m_5\}$, $\{m_7\}$, $\{m_8\}$.
\vspace{-0.2cm}
\item at $p_3$:
$\{m_3,m_1,m_2\}$, $\{m_6,m_4,m_5\}$, $\{m_7\}$, $\{m_8\}$.  
\end{itemize}
Differently, due to the scd-deliveries of the sets including $m_2$ and $m_3$, 
the following  scd-deliveries by $p_1$ and $p_2$ do not satisfy  the
MS-broadcast property: 
\begin{itemize}
\vspace{-0.2cm}
\item at $p_1$:
$\{m_1,m_2\}$, $\{m_3,m_4,m_5\}$, ...  
\vspace{-0.2cm}
\item at $p_2$:
$\{m_1,m_3\}$, $\{m_2\}$, ...
\end{itemize}


\paragraph{A containment property}
\label{sec:properties-SCD}
Let $ms_i^\ell$ be the  $\ell$-the message set  scd-delivered by $p_i$.
Hence, at some time, $p_i$ scd-delivered the sequence
of message sets $ms_i^1,~\ldots, ms_i^x$.
Let $\MS_i^x= ms_i^1\cup \ldots \cup ms_i^x$. 
The following  property follows directly from the 
MS-Ordering and Termination-2 properties: 
\begin{itemize}
\vspace{-0.2cm}
\item Containment.
$\forall~i,j,x,y$:
$(\MS_i^x \subseteq \MS_j^y) \vee (\MS_j^y\subseteq  \MS_i^x)$.
\end{itemize}

\paragraph{Partial order on messages created by the  message sets}
\label{sec:relation-mapsto}
The MS-Ordering and Integrity properties establish a partial order on
the set of all the messages, defined as follows.  Let $\mapsto_i$ be
the local message delivery order at process $p_i$ defined as follows:
$m \mapsto_i m'$ if $p_i$ scd-delivers the message set containing $m$
before the message set containing $m'$.  As no message is
scd-delivered twice, it is easy to see that $ \mapsto_i$ is a partial
order (locally know by $p_i$).
The reader can check that there is a total order (which remains
unknown to the processes)  on the whole set of messages,
that complies with the  partial order $\mapsto = \cup_{1\leq i\leq n}\mapsto_i$. 
This is where \SCD-broadcast can be seen as a weakening
of total order broadcast.

\section{An Implementation of SCD-broadcast}
\label{sec:SCD-implementation}

This section shows that the \SCD-broadcast communication abstraction
is not an oracle-like object (oracles  allow us to extend our
understanding of computing, but cannot be implemented). It describes an
implementation of \SCD-broadcast in an asynchronous send/receive
message-passing system in which any minority of processes may crash.
This system model is denoted  $\CAMP_{n,t}[t<n/2]$ (where 
$\CAMP_{n,t}$ stands for ``Crash Asynchronous Message-Passing'' 
and $t<n/2$ is its restriction  on failures). 
As $t<n/2$ is the  weakest assumption on process failures
that allows a  read/write register to be built on top of an asynchronous 
message-passing system~\cite{ABD95}\footnote{From the point of view of the
  maximal number of process crashes that can be tolerated, assuming
  failures are independent.}, and SCD-broadcast and read/write registers
are computationally equivalent (as shown in the paper), 
the proposed implementation is optimal from a resilience point of view.

\subsection{Underlying communication network}

\paragraph{Send/receive asynchronous network}
Each pair of processes communicate by sending and receiving messages
through two uni-directional channels, one in each direction. Hence,
the communication network is a complete network: any process $p_i$ can
directly send a message to any process $p_j$ (including itself).
A process $p_i$ invokes the operation ``${\sf send}$ {\sc type}($m$)
${\sf to}$ $p_j$'' to send to $p_j$ the message $m$, whose type is
{\sc type}.  The operation ``${\sf receive}$ {\sc type}() ${\sf from}$
$p_j$'' allows $p_i$ to receive from $p_j$ a message whose type is
{\sc type}.

Each channel is reliable (no loss, corruption, nor creation of messages),
not necessarily first-in/first-out, and asynchronous (while the transit time 
of each message is finite, there is no upper bound on message transit times)
Let us notice that, due to process and message asynchrony, no process can
know if another process crashed or is only very slow.

\paragraph{Uniform FIFO-broadcast abstraction}
To simplify the presentation, and without loss of generality, we
consider that the system is equipped with a FIFO-broadcast
abstraction.  Such an abstraction can be built on top of the previous
basic system model without enriching it without additional assumptions
(see e.g.~\cite{R10}).  It is defined by the operations
$\fifobroadcast()$ and $\fifodeliver()$, which satisfy the properties
of Uniform Reliable Broadcast (Validity, Integrity, Termination 1, and
Termination 2), plus the following message ordering property.

\begin{itemize}
\vspace{-0.2cm}
\item FIFO-Order.
  For any pair of processes $p_i$and $p_j$, if $p_i$
  fifo-delivers first  a message $m$ and later a message $m'$,
  both from $p_j$,   no process fifo-delivers $m'$ before $m$.   
\end{itemize}

\subsection{Algorithm}
This section describes Algorithm~\ref{algo:SCD-broadcast-in-MP}, which
implements SCD-broadcast in $\CAMP_{n,t}[t<n/2]$. 
From a terminology point of view, an {\it application message} is a message
that has been scd-broadcast by a process, while a {\it protocol message}
is an implementation message generated by the algorithm.

\paragraph{Local variables at a process $p_i$}
Each process $p_i$ manages the following local
variables. 
\begin{itemize}     
\vspace{-0.1cm}
\item $\bufferi$: buffer (initialized empty)
  where are stored quadruplets containing  messages
  that have been fifo-delivered but not yet scd-delivered in a message set. 
\vspace{-0.2cm}
\item $\todeliveri$: set of quadruplets  containing
  messages to be scd-delivered.
\vspace{-0.2cm}
\item $sn_i$: local logical clock (initialized to $0$),
  which increases by step $1$ and measures the local progress of $p_i$.
  Each application message scd-broadcast by $p_i$ is identified
  by a pair $\langle i,sn\rangle$, where $sn$ is the current value of $sn_i$.  
\vspace{-0.2cm}
\item $clock_i[1..n]$: array of logical dates;
  $clock_i[j]$ is the greatest date $x$ such that the  application
  message $m$  identified $\langle x,j\rangle$ has been
  scd-delivered  by $p_i$.  
\end{itemize}

\paragraph{Content of quadruplet} 
The fields  of a quadruplet
$qdplt=\langle qdplt.msg, qdplt.sd,qdplt.f,qdplt.cl\rangle$
have the following meaning.
\begin{itemize}
\vspace{-0.2cm}
\item $qdplt.msg$ contains an application message $m$,
\vspace{-0.2cm}
\item 
$qdplt.sd$ contains the id of the sender of this application message,
\vspace{-0.2cm}
\item
$qdplt.sn$ contains the local date (seq. number) associated with $m$ by
its sender.  Hence, the pair $\langle qdplt.sd,qdplt.sn\rangle$ is the
identity of $m$.
\vspace{-0.2cm}
\item $qdplt.cl$ is an array of size $n$,
initialized to $[+\infty, \ldots,+\infty]$. 
Then, $qdplt.cl[x]$ will contain the sequence number 
associated with $m$ by $p_x$ when it broadcast \FORWARD$(msg.m,-,-,-,-)$.
This last field is crucial in the scd-delivery
by the process $p_i$ of a message set containing $m$.
\end{itemize}

\paragraph{Protocol message}
The algorithm uses a single type of protocol message denoted
\FORWARD(). Such a message is made up of five fields: an associated
application message $m$, and two pairs, each made up of a sequence
number and a process identity.  The first pair $\langle sd,sn\rangle$ is the
identity of the application message, while the second pair
$\langle f,\snf\rangle$
is the local progress (as captured by $\snf$) of the forwarder
process ${\mathit{p_f}}$
when it forwarded this protocol message to the other processes
by invoking $\fifobroadcast$~\FORWARD$(m,sd,\snsd,p_f,\snf)$
(line~\ref{SCD-from-MP-11}).

\paragraph{Operation $\scdbroadcast()$}
When $p_i$ invokes $\scdbroadcast(m)$, where $m$ is an application message,
it sends the protocol message \FORWARD$(m,i,sn_i,i,sn_i)$ to itself
(this simplifies the writing of the algorithm), and waits until
it has no more message from itself pending in $\buffer_i$, which means
it has scd-delivered a set containing $m$.

\paragraph{Uniform fifo-broadcast of a message \FORWARD}
When a process $p_i$ fifo-delivers a protocol message
\FORWARD$(m,sd,\snsd,f,\snf)$, it first invokes the internal operation
$\forward(m,sd,\snsd,f,\snf)$.  In addition to other statements, the
first fifo-delivery of such a message by a process $p_i$ entails its
participation in the uniform reliable fifo-broadcast of this message
(lines~\ref{SCD-from-MP-05} and~\ref{SCD-from-MP-11}).  In addition to
the invocation of $\forward()$, the fifo-delivery of \FORWARD$()$
invokes also $\trydeliver()$, which strives to scd-deliver a message
set (lines~\ref{SCD-from-MP-04}).

\begin{algorithm}[h!]
\centering{\fbox{
\begin{minipage}[t]{150mm}
\footnotesize 
\renewcommand{\baselinestretch}{2.5}
\resetline
\begin{tabbing}
aaa\=aa\=aaa\=aaaaaa\=\kill

{\bf operation} $\scdbroadcast(m)$ {\bf is}\\

\line{SCD-from-MP-01}  \>\>
$\send$ \FORWARD$(m,sn_i,i, sn_i,i)$ ${\sf to}$ itself;\\

\line{SCD-from-MP-02}  \>\>
    $\wait (\nexists ~qdplt\in \bufferi : qdplt.sd = i)$.\\~\\


{\bf when the message} \= \FORWARD$(m,sd,\snsd,f,\snf)$
          {\bf is fifo-delivered} {\bf do}   $~~$ \% from $\mathit{p_f}$ \\

\line{SCD-from-MP-03}  \>\> $\forward(m,sd,\snsd,f,\snf)$; \\
\line{SCD-from-MP-04}  \>\> $\trydeliver()$.\\~\\


{\bf procedure} $\forward(m,sd,\snsd,f,\snf)$  {\bf is}\\

\line{SCD-from-MP-05} \>\> {\bf if} \=   $(\snsd>clock_i[sd])$\\ 

\line{SCD-from-MP-06} \>\>\> {\bf then} \= {\bf if} \=
$(\exists~ qdplt\in \bufferi: qdplt.sd = sd \land qdplt.sn = \snsd)$\\


\line{SCD-from-MP-07} \>\>\>\>\> {\bf then} \=  $qdplt.cl[f] \leftarrow \snf$\\

\line{SCD-from-MP-08} \>\>\>\>\> {\bf else} \>
          $threshold[1..n]  \leftarrow [\infty,\dots,\infty]$;
          $threshold[f]  \leftarrow \snf$;\\

\line{SCD-from-MP-09} \>\>\>\>\>\>
     {\bf let} $qdplt \leftarrow\langle m, sd, \snsd, threshold[1..n]\rangle$;\\
     
\line{SCD-from-MP-10} \>\>\>\>\>\>$\bufferi \leftarrow \bufferi\cup \{qdplt\}$;\\ 
\line{SCD-from-MP-11} \>\>\>\>\>\>
                             $\fifobroadcast$~\FORWARD$(m,sd,\snsd,i,sn_i)$;\\ 
\line{SCD-from-MP-12} \>\>\>\>\>\> $sn_i \leftarrow sn_i + 1$\\ 
\line{SCD-from-MP-13} \>\>\>\>  {\bf end if}\\ 
\line{SCD-from-MP-14} \>\>  {\bf end if}.\\~\\ 

{\bf procedure} $\trydeliver()$  {\bf is}\\

\line{SCD-from-MP-15} \>\> \textbf{let} \= $\todeliveri \leftarrow
\{qdplt \in \bufferi: |\{f : qdplt.cl[f] < \infty\}| > \frac{n}{2}\} $;\\

\line{SCD-from-MP-16} 
\textbf{while} \=
     $(\exists ~qdplt\in \todeliveri, \exists ~qdplt'\in \bufferi\setminus
     \todeliveri : |\{f : qdplt.cl[f] < qdplt'.cl[f] \}| \le \frac{n}{2}$)
     \textbf{do}\\
  
     \>\>\> $\todeliveri\leftarrow \todeliveri\setminus \{qdplt\}$
     \textbf{end while};\\

\line{SCD-from-MP-17} \>\> {\bf if} \= $(\todeliveri \neq \emptyset)$\\

\line{SCD-from-MP-18} \>\>\> {\bf then} \=
     {\bf for each} $qdplt\in \todeliveri$
     {\bf do} $clock_i[qdplt.sd] \leftarrow \max(clock_i[qdplt.sd],qdplt.sn)$ 
     {\bf end for}; \\
    \line{SCD-from-MP-19} \>\>\>\>
                         $\bufferi\leftarrow \bufferi\setminus \todeliveri$;\\
\line{SCD-from-MP-20} \>\>\>\>
     $ms \leftarrow\{m : \exists~ qdplt\in \todeliveri : qdplt.msg = m\}$;
    $\scdeliver(ms)$\\

\line{SCD-from-MP-21} \>\> {\bf end if}.

\end{tabbing}
\end{minipage}
}
\caption{An implementation of
         \SCD-broadcast in  $\CAMP_{n,t}[t<n/2]$ (code for $p_i$)}
\label{algo:SCD-broadcast-in-MP}
}
\end{algorithm}

\paragraph{The core of  the algorithm}
Expressed with the relations $\mapsto_i$, $1 \leq i\leq n$, introduced in
Section~\ref{sec:relation-mapsto}, the main issue of the algorithm is
to ensure that, if there are two message $m$ and $m'$ and a process
$p_i$ such that $m\mapsto_i m'$, then there is no $p_j$ such that
$m'\mapsto_j m$.

To this end, a process $p_i$ is allowed to scd-deliver a message $m$
before a message $m'$ only if it knows that a majority of processes
$p_j$ have fifo-delivered a message \FORWARD$(m,-,-,-)$ before $m'$;
$p_i$ knows it (i) because it fifo-delivered from $p_j$ a message
\FORWARD$(m,-,-,-,-)$ but not yet a message
\FORWARD$(m',-,-,-,-)$, or (ii) because it fifo-delivered from $p_j$
both the messages \FORWARD$(m,-,-,-,snm)$ and
\FORWARD$(m',-,-,-,snm')$ and the sending date $smn$ is smaller than
the sending date $snm'$.  The MS-Ordering property follows then from
the impossibility that a majority of processes ``sees $m$ before $m'$'',
while another majority ``sees $m'$ before $m$''.

\paragraph{Internal operation $\forward()$}
This operation can be seen as an enrichment (with the fields $f$ and $\snf$)
of the reliable fifo-broadcast implemented by the messages
\FORWARD$(m,sd,\snsd,-,-)$.
Considering such a message \FORWARD$(m,sd,\snsd,f,\snf)$, 
$m$ was scd-broadcast by $p_{sd}$ at its local time $\snsd$, and
relayed by the forwarding process  $p_f$ at its local time $\snf$.
If $\snsd \leq  clock_i[sd]$, $p_i$ has already  scd-delivered a message set
containing $m$ (see lines~\ref{SCD-from-MP-18} and \ref{SCD-from-MP-20}). 
If $\snsd > clock_i[sd]$, there are two cases
defined by the predicate of line~\ref{SCD-from-MP-06}.
\begin{itemize}
\vspace{-0.1cm}
\item There is no quadruplet $qdplt$ in $\bufferi$ such that
  $qdplt.msg=m$.  In this case, $p_i$ creates a quadruplet
  associated with $m$, and adds
  it to $\bufferi$ (lines~\ref{SCD-from-MP-08}-\ref{SCD-from-MP-10}).
  Then, $p_i$ participates in the fifo-broadcast of $m$
  (line~\ref{SCD-from-MP-11}) and  records its local  progress by
  increasing $sn_i$ (line~\ref{SCD-from-MP-12}).
\vspace{-0.2cm}
\item
There is a quadruplet  $qdplt$ in $\bufferi$ 
associated with $m$, i.e., $qdplt=\langle m,-,-, -\rangle \in \bufferi$.
In this case,  $p_i$  assigns $\snf$
to $qdplt.cl[f]$ (line~\ref{SCD-from-MP-07}), thereby indicating that
$m$ was known and forwarded by $\mathit{p_f}$ at its local time $\snf$.
\end{itemize}

\paragraph{Internal operation $\trydeliver()$}
When it executes $\trydeliver()$, $p_i$ first computes  the set
$\todeliveri$ of the quadruplets $qdplt$ containing  application
messages $m$ which have been seen by a majority of processes
(line~\ref{SCD-from-MP-15}).  From $p_i$'s point of view, a message has been
seen by a process $\mathit{p_f}$ if $qdplt.cl[f]$ has been set to a
finite value (line~\ref{SCD-from-MP-07}).

As indicated in a previous paragraph, if a majority of processes
received first a message \FORWARD\ carrying $m'$ and later another
message \FORWARD\ carrying $m$, it might be that some process $p_j$
scd-delivered a set containing $m'$ before scd-delivering a set
containing $m$.  Therefore, $p_i$ must avoid scd-delivering a set
containing $m$ before scd-delivering a set containing $m'$.  This is
done at line~\ref{SCD-from-MP-16}, where $p_i$ withdraws the
quadruplet $qdplt$ corresponding to $m$ if it has not enough
information to deliver $m'$ (i.e. the corresponding $qdplt'$ is not in
$\todeliver_i$) or it does not have the proof that the situation
cannot happen, i.e. no majority of processes saw the message
corresponding to $qdplt$ before the message corresponding to $qdplt'$
(this is captured by the predicate
$|\{f : qdplt.cl[f] < qdplt'.cl[f] \}| \le \frac{n}{2}$).

If $\todeliveri$ is not empty after it has been purged
(lines~\ref{SCD-from-MP-16}-\ref{SCD-from-MP-17}), $p_i$ computes a
message set to scd-deliver.  This set $ms$ contains all the application
messages in the quadruplets of $\todeliveri$
(line~\ref{SCD-from-MP-20}).  These quadruplets are withdrawn from
$\bufferi$ (line~\ref{SCD-from-MP-18}).  Moreover, before this
scd-delivery, $p_i$ needs to updates $clock_i[x]$ for all the entries
such that $x=qdplt.sd$ where $qdplt\in \todeliveri$
(line~\ref{SCD-from-MP-18}).  This update is needed to ensure that
the future uses of the predicate of line~\ref{SCD-from-MP-17} are correct.

\subsection{Cost and proof of correctness}

\begin{lemma}
\label{lemma-broadcast-validity}
If a process  scd-delivers a message set containing  $m$, some
process invoked $\scdbroadcast(m)$.
\end{lemma}

\begin{proofL}
  If a process $p_i$ scd-delivers a set containing a message $m$, it
  previously added into $\bufferi$ a quadruplet $qdplt$
  such that $qdplt.msg=m$ (line \ref{SCD-from-MP-10}), for which it
  follows that it fifo-delivered a  protocol message \FORWARD$(m,-,-,-,-)$.
  Due to the fifo-validity property, it follows that a process
  generated the fifo-broadcast of this message, which
  originated from an invocation of $\scdbroadcast(m)$.
  \renewcommand{\toto}{lemma-broadcast-validity}
\end{proofL}

\begin{lemma}
\label{lemma-broadcast-integrity}
No process scd-delivers  the same message twice. 
\end{lemma}

\begin{proofL}
 Let us observe that, due to the wait statement at
 line~\ref{SCD-from-MP-02}, and the increase of $sn_i$ at
 line~\ref{SCD-from-MP-15} between two successive scd-broadcast by a
 process $p_i$, no two application messages can have the same identity
 $\langle i,sn\rangle$. It follows that there is a single quadruplet
 $\langle m, i,sn, -\rangle$ that can be added to $\bufferi$, and this
 is done only once (line~\ref{SCD-from-MP-10}).  Finally, let us
 observe that this quadruplet is suppressed from $\bufferi$, just
 before $m$ is scd-delivered (line~\ref{SCD-from-MP-19}-\ref{SCD-from-MP-20}),
 which concludes the proof of the lemma.
 \renewcommand{\toto}{lemma-broadcast-integrity}
\end{proofL}

\begin{lemma}
\label{lemma:broadcast}
If a process $p_i$ executes  $\fifobroadcast$~\FORWARD$(m,sd,\snsd,i,sn_i)$
(i.e., executes line~{\em\ref{SCD-from-MP-19}}), each non-faulty process $p_j$
executes once $\fifobroadcast$~\FORWARD$(m, sd, \snsd,j, sn_j)$.
\end{lemma}

\begin{proofL}
  First, we prove that $p_j$ broadcasts a message \FORWARD$(m,
  sd,\snsd,j, sn_j)$.  As $p_i$ is non-faulty, $p_j$ will eventually
  receive the message sent by $p_i$. At that time, if $\snsd >
  clock_j[sd]$, after the condition on line~\ref{SCD-from-MP-06} and
  whatever its result, $\buffer_i$ contains a quadruplet $qdplt$ with
  $qdplt.sd = sd$ and $qdplt.sn =\snsd$.  That $qdplt$ was inserted at
  line~\ref{SCD-from-MP-10} (possibly after the reception of a
  different message), just before $p_j$ sent a message
  \FORWARD$(m, sd, \snsd, j,sn_j)$ at line~\ref{SCD-from-MP-11}.  Otherwise,
  $clock_j[sd]$ was incremented on line~\ref{SCD-from-MP-18}, when
  validating some $qdplt'$ added to $\bufferj$ after $p_j$ received a
  (first) message \FORWARD$(qdplt'.msg,sd,\snsd,f, clock_f[sd])$ from
  $p_f$. Because the messages \FORWARD$()$ are fifo-broadcast (hence
  they are delivered in their sending order), $p_{sd}$ sent message
  \FORWARD$(qdplt.msg, sd, \snsd,sd, \snsd)$ before
  \FORWARD$(qdplt'.msg, sd, clock_j[sd], sd,clock_j[sd])$, and all
  other processes only forward messages, $p_j$ received 
  \FORWARD$(qdplt.msg, sd, \snsd,-,-)$ from $p_f$ before the message
  \FORWARD$(qdplt'.msg, sd, clock_j[sd],-,-)$.  At that time, $\snsd >
  clock_j[sd]$, so the previous case applies.

After $p_j$ broadcasts its message \FORWARD$(m, sd, \snsd,j, sn_j)$ on
line~\ref{SCD-from-MP-11}, there is a $qdplt\in \buffer_j$ with $ts(qdplt)
= \langle sd, \snsd \rangle$, until it is removed on
line~\ref{SCD-from-MP-16} and $clock_j[sd] \ge \snsd$. Therefore, one of the
conditions at lines~\ref{SCD-from-MP-05} and~\ref{SCD-from-MP-06} will
stay false for the stamp $ts(qdplt)$ and $p_j$ will never execute
line~\ref{SCD-from-MP-11} with the same stamp $\langle sd, \snsd\rangle$ later.
\renewcommand{\toto}{lemma:broadcast}
\end{proofL}

\begin{lemma}
  \label{lemma-broadcast-sc-ordering}
  Let $p_i$ be a process that  scd-delivers a set $ms_i$ containing a
  message $m$ and later  scd-delivers a set $ms'_i$ containing a message
  $m'$.  No process $p_j$  scd-delivers first a set $ms'_j$ containing
  $m'$  and later a message set $ms_j$ containing $m$.
\end{lemma}

\begin{proofL}
  Let us suppose there are two messages $m$ and $m'$ and two processes
  $p_i$ and $p_j$ such that $p_i$ scd-delivers a set $ms_i$
  containing $m$ and later scd-delivers a set $ms'_i$ containing $m'$
  and $p_j$ scd-delivers a set $ms'_j$ containing $m'$ and later
  scd-delivers a set $ms_j$ containing $m$.

  When $m$ is delivered by $p_i$, there is an element $qdplt\in
  \buffer_i$ such that $qdplt.msg = m$ and because of line
  \ref{SCD-from-MP-15}, $p_i$ has received a message \FORWARD$(m,-,-,-,-)$
  from more than $\frac{n}{2}$ processes.
  \begin{itemize}
  \vspace{-0.2cm}
  \item If there is no element $qdplt'\in \buffer_i$ such that $qdplt'.msg =
    m'$, since $m'$ has not been delivered by $p_i$ yet, $p_i$ has not
    received a message \FORWARD$(m',-,-,-,-)$ from any process
    (lines \ref{SCD-from-MP-10} and \ref{SCD-from-MP-19}). Hence,
    because the communication channels are FIFO, more than
    $\frac{n}{2}$ processes have sent a message \FORWARD$(m,-,-,-,-)$
    before sending a message \FORWARD$(m',-,-,-,-)$.
    \vspace{-0.2cm}
   \item Otherwise, $qdplt'\notin \todeliver_i$ after
    line~\ref{SCD-from-MP-16}. As the communication channels are
    FIFO, more than half of the processes have sent a message
    \FORWARD$(m,-,-,-,-)$ before a message \FORWARD$(m',-,-,-,-)$.
  \end{itemize}

  Using the same reasoning, it follows that
  when $m'$ is delivered by $p_j$, more than
  $\frac{n}{2}$ processes have sent a message \FORWARD$(m',-,-,-,-)$ before
  sending a message \FORWARD$(m,-,-,-,-)$.  There exists a process $p_k$ in
  the intersection of the two majorities, that has (a) sent 
  \FORWARD$(m,-,-,-,-)$ before sending \FORWARD$(m',-,-,-,-)$ and
  (b) sent 
  \FORWARD$(m',-,-,-,-)$ before sending a message \FORWARD$(m,-,-,-,-)$.
  However, it follows from
  Lemma~\ref{lemma:broadcast} that  $p_k$ can  send a single message
  \FORWARD$(m',-,-,-,-)$ and a single
  message \FORWARD$(m,-,-,-,-)$, which leads to a  contradiction.
\renewcommand{\toto}{lemma-broadcast-sc-ordering}
\end{proofL}

\begin{figure}[h!]
  \begin{center}
    \begin{tikzpicture}

      \draw[->] (0,0) node[left]{$p_i$} -- (11,0);
      \draw[->] (0,1.5) node[left]{$p_f$} -- (11,1.5);

      \draw (2.75,2.1) node{\footnotesize$\scdbroadcast(m_k)$};
      \draw (3.4,0.75) node[left]{\footnotesize\FORWARD$(m_k,f,sn_f(k),f,sn_f(k))$};
      \draw (10.2,0.75) node[right]{$\cdots$};

      \draw[fill=white] (2,1.2) rectangle (3.5,1.8);
      \draw (2.1,1.5) node{$\bullet$};
      \draw[-latex] (2.3,1.5) to[out=0,in=120,distance=20] (3.8,0.1);
      \draw[-latex] (2.2,1.6) to[out=60,in=120,distance=5] (2.5,1.6);

      \draw[fill=white] (4.8,1.2) rectangle (7.5,1.8);
      \draw (4.9,1.5) node{$\bullet$};
      \draw[-latex] (5,1.4) -- (6,0.1);
      \draw[-latex] (5,1.6) to[out=60,in=120,distance=5] (5.3,1.6);

      \draw[fill=white] (7.8,1.2) rectangle (9.5,1.8);
      \draw (7.9,1.5) node{$\bullet$};
      \draw[-latex] (8,1.4) -- (9,0.1);
      \draw[-latex] (8,1.6) to[out=60,in=120,distance=5] (8.3,1.6);

      \draw (5.5,0.75) node[right]{\footnotesize$sn_f(k1)$};
      \draw (8.5,0.75) node[right]{\footnotesize$sn_f(k2)$};

      \draw[->] (4.8,-0.3) -- (5.8,-0.3);
      \draw (5.9,-0.3) node{${}_i^\star$};
      \draw[->] (6.1,-0.3) -- (8.8,-0.3);
      \draw (8.9,-0.3) node{${}_i^\star$};
      \draw[->] (8.8,-0.6) -- (4.8,-0.6);
      \draw (4.7,-0.6) node{${}_i^\star$};

      \draw (1.5,-0.9) node[below]{\footnotesize\FORWARD$(m,sd,\snsd,-,-)$};
      \draw (4,-1.3) node[below]{\footnotesize\FORWARD$(m,sd,\snsd,-,-)$};
      \draw[-latex,dashed,thick] (1.5,-0.9) -- (1.8,-0.1);
      \draw[-latex,dashed,thick] (4,-1.3) -- (4.2,1.4);
      \draw[-latex,dashed,thick] (4.3,1.4) -- (4.7,0.1);
      \draw[-latex,dashed,thick] (4.3,1.6) to[out=60,in=120,distance=5] (4.6,1.6);

      \draw (7.1,2.5) node[above]{\footnotesize\FORWARD$(m_{l+1},sd_{l+1},sn_{sd_{l+1}},-,-)$};
      \draw[-latex,dotted,thick] (7.0,2.5) -- (7.2,1.55);
      \draw[-latex,dotted,thick] (7.1,2.5) -- (7.3,1.55);
      \draw[-latex,dotted,thick] (7.2,2.5) -- (7.4,1.55);
      \draw[-latex,dotted,thick] (7.2,1.45) -- (7.4,0.05);
      \draw[-latex,dotted,thick] (7.3,1.45) -- (7.5,0.05);
      \draw[-latex,dotted,thick] (7.4,1.45) -- (7.6,0.05);

    \end{tikzpicture}
  \end{center}
  \caption{Message pattern introduced in Lemma~\ref{lemma:liveness}}
  \label{fig:lemma:liveness}
\end{figure}
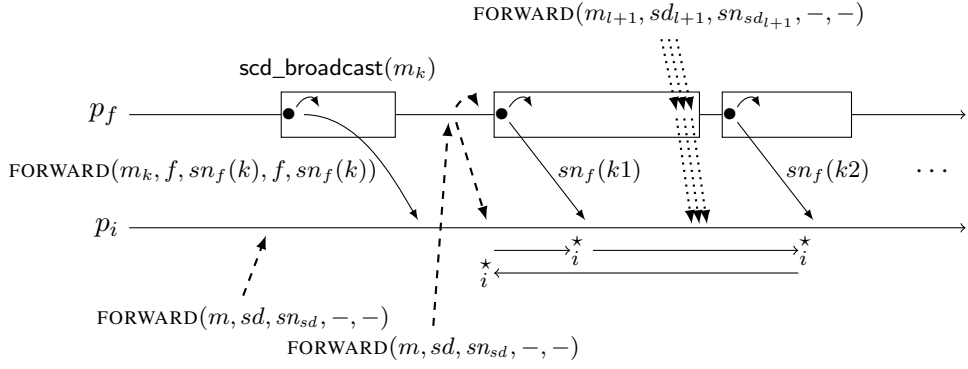

\begin{lemma} \label{lemma:liveness}
  If a non-faulty process  executes
  $\fifobroadcast$~\FORWARD$(m,sd,\snsd,i,sn_i)$
  (line~{\em\ref{SCD-from-MP-11}}), 
  it  scd-delivers a message set containing $m$.
\end{lemma}

\begin{proofL}
  Let $p_i$ be a non-faulty process.   For any pair of messages
  $qdplt$ and $qdplt'$ ever inserted
  in $\bufferi$,  let $ts = ts(qdplt)$ and  $ts' = ts(qdplt')$.
  Let  $\rightarrow_i$ be the dependency relation defined  as follows:
  $ts \rightarrow_i ts' \stackrel{def}{=}|\{j : qdplt'.cl[j] < qdplt.cl[j]
  \}| \le \frac{n}{2}$ (i.e. the dependency does not exist if $p_i$
  knows that a majority of processes have seen the first update
  --due to $qdplt'$-- before the second --due to $qdplt$).
  Let $\rightarrow_i^\star$ denote the transitive  closure of $\rightarrow_i$.
  
  Let us suppose (by contradiction) that the timestamp $\langle sd,
  \snsd\rangle$ associated with the message $m$ (carried by the
  protocol message \FORWARD$(m, sd, \snsd,i, sn_i)$ fifo-broadcast by
  $p_i$), has an infinity of predecessors according to
  $\rightarrow_i^\star$.  As the number of processes is finite, an
  infinity of these predecessors have been generated by the same
  process, let us say $p_f$.  Let $\langle f, sn_f(k) \rangle_{k\in
    \mathbb{N}}$ be the infinite sequence of the timestamps associated
  with the invocations of the $\scdbroadcast()$ issued by $p_f$. The
  situation is depicted by Figure~\ref{fig:lemma:liveness}.
  
  As $p_i$ is non-faulty, $p_f$ eventually receives a message
  \FORWARD$(m,sd, \snsd, i, sn_i)$, which means $p_f$ broadcast an
  infinity of messages \FORWARD$(m(k), f, sn_f(k), f, sn_f(k))$ after
  having broadcast the message 
  \FORWARD$(m, sd, \snsd, f, sn_f)$. Let $\langle f, sn_f(k1) \rangle$
  and $\langle f, sn_f(k2) \rangle$ be the timestamps associated with the
  next two messages sent by $p_f$, with $sn_f(k1) < sn_f(k2)$.  By
  hypothesis, we have $\langle f, sn_f(k2) \rangle\rightarrow_i^\star
  \langle sd, \snsd \rangle$.  Moreover, all processes received their
  first message \FORWARD$(m, sd, \snsd, -,-)$ before their first
  message \FORWARD$(m(k), f, sn_f(k), -,-)$, so $\langle sd, \snsd
  \rangle\rightarrow_i^\star \langle f, sn_f(k1) \rangle$. Let us
  express the path $\langle f, sn_f(k2) \rangle\rightarrow_i^\star
  \langle f, sn_f(k1) \rangle$:\\ $\langle f, sn_f(k2) \rangle =
  \langle sd'(1), sn'(1) \rangle \rightarrow_i \langle sd'(2), sn'(2)
  \rangle \rightarrow_i \dots \rightarrow_i \langle sd(m), sn'(m)
  \rangle = \langle f, sn_f(k1) \rangle$.

  In the time interval starting when $p_f$ sent the message
  \FORWARD$(m(k1), f, sn_f(k1), f, sn_f(k1))$ and finishing when it
  sent the message \FORWARD$(m(k2), f, sn_f(k2), f, sn_f(k2))$,
  the waiting condition of line~\ref{SCD-from-MP-02} became true, so $p_f$
  scd-delivered a set containing the message $m(k1)$, and according
  to Lemma~\ref{lemma-broadcast-validity}, no set containing the
  message $m(k2)$. Therefore, there is an index $l$ such that process
  $p_f$ delivered sets containing messages associated with a timestamp
  $\langle sd'(l), sn'(l) \rangle$ for all $l'>l$ but not for
  $l'=l$. Because the channels are FIFO and thanks to
  lines~\ref{SCD-from-MP-15} and~\ref{SCD-from-MP-16}, it means that a
  majority of processes have sent a message \FORWARD$(-, sd'(l+1),
  sn'(l+1),-,-)$ before a message \FORWARD$(-, sd'(l), sn'(l),-,-)$,
  which contradicts the fact that $\langle sd'(l), sn'(l) \rangle
  \rightarrow_i \langle sd'(l+1), sn'(l+1) \rangle$.

  Let us suppose a non-faulty process $p_i$ has fifo-broadcast a
  message \FORWARD$(m, sd, \snsd,i, sn_i)$
  (line~\ref{SCD-from-MP-10}).  It inserted a quadruplet $qdplt$ with
  timestamp $\langle sd, \snsd\rangle$ on line~\ref{SCD-from-MP-09}
  and by what precedes, $\langle sd, \snsd\rangle$ has a finite number
  of predecessors $\langle sd_1, sn_1\rangle, \dots, \langle sd_l,
  sn_l\rangle $ according to $\rightarrow_i^\star$.  As $p_i$ is
  non-faulty, according to Lemma~\ref{lemma:broadcast}, it eventually
  receives a message \FORWARD$(-,sd_k, sn_k,-,-)$ for all $1\le k \le
  l$ and from all non-faulty processes, which are in majority.

  Let $\mathit{pred}$ be the set of all quadruplets $qdplt'$ such that
  $\langle qdplt'.sd, qdplt'.sn \rangle \rightarrow_i^\star \langle sd,
  \snsd \rangle$.  Let us consider the moment when $p_i$ receives the
  last message \FORWARD$(-,sd_k, sn_k, f, sn_f)$ sent by a correct
  process $p_f$.  For all $qdplt'\in \mathit{pred}$, either $qdplt'.msg$ has
  already been delivered or $qdplt'$ is inserted $\todeliver_i$ on
  line~\ref{SCD-from-MP-15}.  Moreover, no $qdplt'\in \mathit{pred}$
  will be removed from $\todeliver_i$, on line~\ref{SCD-from-MP-16},
  as the removal condition is the same as the definition of
  $\rightarrow_i$.  In particular for $qdplt' = qdplt$, either $m$ has
  already been scd-delivered or $m$ is present in $\todeliver_i$ on
  line~\ref{SCD-from-MP-17} and will be scd-delivered on
  line~\ref{SCD-from-MP-20}.
\renewcommand{\toto}{lemma:liveness}
\end{proofL}

\begin{lemma}
\label{lemma-broadcast-termination1}
If a non-faulty process scd-broadcasts a message $m$, it scd-delivers
a message set containing $m$.
\end{lemma}

\begin{proofL}
  If a non-faulty process scd-broadcasts a message $m$, it previously 
  fifo-broadcast the message
  \FORWARD$(m, sd, \snsd,i, sn_i)$ at line~\ref{SCD-from-MP-11}).
  Then, due to Lemma~\ref{lemma:liveness}, it scd-delivers a message set
  containing $m$.
  \renewcommand{\toto}{lemma-broadcast-termination1}
\end{proofL}

\begin{lemma}
\label{lemma-broadcast-termination2}
If a process scd-delivers a message $m$, every  non-faulty process
scd-delivers a message set containing $m$.
\end{lemma}

\begin{proofL}
  Let $p_i$ be a process $p_i$ that scd-delivers a message $m$. At
  line~\ref{SCD-from-MP-20},
  there is a quadruplet $qdplt\in \todeliver_i$ such that $qdplt.msg = m$.
  At line~\ref{SCD-from-MP-15}, $qdplt\in \buffer_i$, and 
  $qdplt$ was inserted in $\buffer_i$ at line~\ref{SCD-from-MP-10},
  just before $p_i$ fifo-broadcast the message \FORWARD$(m, sd, \snsd,i, sn_i)$.
  By Lemma~\ref{lemma:broadcast}, every non-faulty process $p_j$ sends
  a message \FORWARD$(m, sd, \snsd,j, sn_j)$, so by Lemma~\ref{lemma:liveness},
  $p_j$ scd-delivers a message set containing $m$.
 \renewcommand{\toto}{lemma-broadcast-termination2}
\end{proofL}

\begin{theorem}
\label{theorem:SCD}
Algorithm~{\em\ref{algo:SCD-broadcast-in-MP}}
implements the {\em \SCD-broadcast} communication abstraction
in $\CAMP_{n,t}[t<n/2]$. Moreover, it requires $O(n^2)$
messages per invocation of $\scdbroadcast()$. If there is an upper bound
$\Delta$ on messages transfer delays (and local computation times are
equal to zero), each {\em \SCD-broadcast} costs at most $2\Delta$ time units.
\end{theorem}

\begin{proofT}
The proof follows from
Lemma~\ref{lemma-broadcast-validity} (Validity), 
Lemma~\ref{lemma-broadcast-integrity} (Integrity), 
Lemma~\ref{lemma-broadcast-sc-ordering} (MS-Ordering), 
Lemma~\ref{lemma-broadcast-termination1} (Termination-1), and
Lemma~\ref{lemma-broadcast-termination2} (Termination-2).

The  $O(n^2)$ message complexity comes from the fact that,
due to the predicates of line~\ref{SCD-from-MP-05} and~\ref{SCD-from-MP-06},
each application message $m$ is forwarded at most once by each process
(line~\ref{SCD-from-MP-11}). The  $2\Delta$ follows from the same argument.
\renewcommand{\toto}{theorem:SCD}    
\end{proofT}

The next corollary follows from (i)
Theorems~\ref{theorem:proof-snapshot}  and~\ref{theorem:SCD}, and (ii) 
the fact that the constraint
$(t<n/2)$ is an upper bound on the number of faulty processes
to build a read/write register (or snapshot object)~\cite{ABD95}.

\begin{corollary}
\label{coro:sc-optimal}
Algorithm~{\em\ref{algo:SCD-broadcast-in-MP}} is resiliency optimal. 
\end{corollary}
  
\section{An SCD-broadcast-based Communication Pattern}
\label{sec:communication-pattern}
All the algorithms implementing concurrent objects and tasks, which
are presented in this paper, are based on the same communication
pattern, denoted Pattern~\ref{communication-pattern}.  This pattern
involves each process, either as a client (when it invokes an
operation), or as a server (when it scd-delivers a message set).

When a process $p_i$ invokes an operation $\op()$, it executes once
the lines~\ref{pattern-01}-\ref{pattern-03} for a task, and $0$, $1$,
or $2$ times for an operation on a concurrent object.  In this last
case, this number of times depends on the consistency condition which is
implemented (linearizability~\cite{HW90} or sequential consistency~\cite{L79}).

\begin{pattern}[h!]
\centering{\fbox{
\begin{minipage}[t]{150mm}
\footnotesize 
\renewcommand{\baselinestretch}{2.5}
\resetline
\begin{tabbing}
aaa\=aa\=aaa\=aaaaaa\=\kill

{\bf operation} $\op()$ {\bf is}\\

\>\>According to the object/task that is implemented, and
its  consistency condition (if it is an object, \\
\>\> linearizability vs seq. consistency),
execute $0$, $1$, or $2$ times the lines
\ref{pattern-01}-\ref{pattern-03} where the  message type \\
\>\> \TYPE\ is either a pure synchronization message \SYNC\
or an object/task-dependent message \MSG; \\
\line{pattern-01}  \>\> $\done_i \leftarrow \ffalse$;\\
\line{pattern-02}  \>\> $\scdbroadcast$ \TYPE $(a,b,...,i)$;\\
\>\>
$a, b, ...$ are data, and $i$ is  the id of the invoking process;
 a message \SYNC\ carries only the id of its sender;\\

\line{pattern-03}  \>\>  $\wait (\done_i)$;\\
\line{pattern-04}
  \>\> According to the states of the local variables, compute a result $r$;
      $\return (r)$. \\~\\

{\bf when the message set} \= 
$\{$~\MSG$(...,j_1),~\ldots,$ \MSG$(...,j_x)),$ 
   \SYNC$(j_{x+1}),~\ldots,$ \SYNC$(j_{y})~\}$ {\bf is scd-delivered do}\\
   \line{pattern-05}
  
\>\> {\bf for} 
     {\bf each message} $m =$ \MSG$(..., j)$ {\bf do}
     statements specific to the object/task that is implemented
     {\bf end for};\\

\line{pattern-06}
\>\> {\bf if} $\exists \ell~:~j_\ell=i$ 
                {\bf then} $\done_i \leftarrow \ttrue$ {\bf end if}. 

\end{tabbing}
\end{minipage}
}
  \caption{Communication pattern (Code for $p_i$)}
    \label{communication-pattern}
}
\end{pattern}

All the messages sent by a process $p_i$ are used to synchronize its
local data representation of the object,  or its local view
of the current state of the task.  This
synchronization is realized by the Boolean $\done_i$ and the parameter
$i$ carried by every message (lines~\ref{pattern-01}, 
\ref{pattern-03}, and \ref{pattern-06}): $p_i$ is blocked until the
message it scd-broadcast just before is scd-delivered.  The values
carried by a message \MSG\ are related to the object/task that is
implemented, and may require local computation.

It appears that the combination of this communication pattern and the
properties of SCD-broadcast provides us with a single simple framework
that allows for correct implementations of both concurrent objects and
tasks.

The next three  sections describe  algorithms implementing a snapshot
object, a counter object, and the lattice agreement task,
respectively. All these algorithms consider the system model
$\CAMP_{n,t}[\emptyset]$ enriched with the SCD-broadcast communication
abstraction, denoted $\CAMP_{n,t}$[\SCD-broadcast], and
use the previous communication pattern.

\section{SCD-broadcast in Action (its Power): Snapshot Object} 
\label{sec:SCD-power-snapshot}

\subsection{Snapshot object}

\paragraph{Definition}
The snapshot  object was introduced in~\cite{AADGMS93,A94}.
A snapshot object is an array $\REG[1..m]$ of atomic read/write registers
which provides the processes with two operations,
denoted $\wwrite(r,-)$ and $\snapshot()$.
The invocation of $\wwrite(r,v)$, where $1\leq r\leq m$,
by a process $p_i$  assigns atomically $v$ to $\REG[r]$.
The invocation of  $\snapshot()$ returns the value of $\REG[1..m]$
as if it was  executed instantaneously.  Hence, in any execution of a snapshot
object, its operations  $\wwrite()$ and $\snapshot()$ are linearizable.

The underlying atomic registers can be Single-Reader (SR) or
Multi-Reader (MR) and Single-Writer (SR) or Multi-Writer (MW).  We
consider only SWMR and MWMR registers.  If the registers are SWMR the
snapshot is called SWMR snapshot (and we have then $m=n$).  Moreover,
we always have $r=i$, when $p_i$ invokes $\wwrite(r,-)$.  If the
registers are MWMR, the snapshot object is called MWMR.

\paragraph{Implementations based on  read/write  registers}
\label{sec:rw-snapshot}
Implementations of both  SWMR and  MWMR snapshot objects on top of
read/write atomic registers have been proposed
(e.g.,~\cite{AADGMS93,A94,IR12,ICMT94}). 
The ``hardness'' to build snapshot objects in read/write systems and
associated lower bounds are presented in the  survey~\cite{E05}.
The best algorithm known to implement  an SWMR snapshot requires 
$O(n \log n)$ read/write on the base SWMR registers  for both 
the $\wwrite()$ and $\snapshot()$ operations~\cite{AR98}.
As far as MWMR snapshot objects are concerned, there are implementations 
where each operation has an $O(n)$ cost\footnote{Snapshot objects built
  in read/write  models enriched with operations such as Compare\&Swap,
  or LL/SC, have also been considered, e.g.,\cite{J05,IR12}. Here
  we are interested in pure read/write models.}. 

As far as the construction of an SWMR (or MWMR) snapshot object in 
crash-prone asynchronous message-passing systems where $t<n/2$ is
concerned, it is possible to stack two constructions: first 
an algorithm implementing SWMR (or MWMR) atomic read/write registers
(e.g.,~\cite{ABD95})), and, on top of it, an algorithm implementing an
SWMR (or MWMR) snapshot object. This stacking approach provides
objects whose operation cost is $O(n^2 \log n)$ messages for SWMR
snapshot, and $O(n^2)$ messages for MWMR snapshot.  An algorithm
based on the same low level communication pattern
as  the one used in~\cite{ABD95}, which builds an atomic
SWMR snapshot object ``directly'' (i.e., without stacking algorithms)
was recently presented in~\cite{DFRR16} (the aim of this algorithm is to
perform better that the stacking approach in concurrency-free executions). 

\subsection{An algorithm for atomic  MWMR snapshot 
  in $\CAMP_{n,t}$[\SCD-broadcast]}

\paragraph{Local representation of  $\REG$ at a process $p_i$}
At each register $p_i$,  $\REG[1..m]$ is represented by three local variables
$reg_i[1..m]$ (data part), plus  $tsa_i[1..m]$ and $done_i$ (control part). 
\begin{itemize}
\vspace{-0.1cm}
\item $\done_i$ is a Boolean variable. 
\vspace{-0.2cm}
\item $\reg_i[1..m] $ contains the current value of  $\REG[1..m]$,
  as known by $p_i$.
\vspace{-0.2cm}
\item $tsa_i[1..m]$ is an array of timestamps associated with the values
  stored in $\reg_i[1..m]$.
A timestamp is a pair made of a local clock value and a process identity. 
Its initial value is $\langle 0,-\rangle$.
The fields associated with $tsa_i[r]$ are denoted
 $\langle tsa_i[r].date ,tsa_i[r].proc \rangle$. 
\end{itemize}

\paragraph{Timestamp-based order relation}
We consider the classical lexicographical total order relation
on timestamps, denoted  $<_{ts}$. Let $ts1= \langle h1,i1\rangle$ and
$ts2= \langle h2,i2\rangle$. We have 
$ ts1 \lexts ts2 \stackrel{\mathit{def}}{=}
(h1<h2) \vee ((h1=h2)\wedge (i1<i2)).$

\paragraph{Algorithm~\ref{algo:snapshot-from-SCD}: snapshot operation}
This algorithm consists of one instance of the communication pattern
introduced in Section~\ref{sec:communication-pattern}
(line~\ref{Snap-from-SC-01}), followed by the return of the local  value
of $\reg_i[1..m]$ (line~\ref{Snap-from-SC-02}).
The message \SYNC$(i)$, which is  scd-broadcast is a pure
synchronization message,
whose aim is to entail the refreshment of the value of $\reg_i[1..m]$
(lines~\ref{Snap-from-SC-05}-\ref{Snap-from-SC-11}) which occurs before
the setting of $\done_i$ to $\ttrue$ (line~\ref{Snap-from-SC-12}).

\begin{algorithm}[h!]
\centering{\fbox{
\begin{minipage}[t]{150mm}
\footnotesize 
\renewcommand{\baselinestretch}{2.5}
\resetline
\begin{tabbing}
aaa\=aa\=aaa\=aaaaaa\=\kill

{\bf operation} $\snapshot()$ {\bf is}\\

\line{Snap-from-SC-01}  \>\> $\done_i \leftarrow \ffalse$;
$\scdbroadcast$ \SYNC $(i)$; $\wait (\done_i)$;\\

\line{Snap-from-SC-02}  \>\> $\return (\reg_i[1..m])$. \\~\\

{\bf operation} $\wwrite(r, v)$ {\bf is}\\

\line{Snap-from-SC-03}  \>\> $\done_i \leftarrow \ffalse$;
$\scdbroadcast$ \SYNC $(i)$; $\wait (\done_i)$;\\

\line{Snap-from-SC-04}  \>\>
$\done_i \leftarrow \ffalse$;
$\scdbroadcast$\WRITE $(r,v, \langle \tsa_i[r].date +1, i \rangle)$; 
$\wait (\done_i)$.~\\~\\

{\bf when the message set} \= 
$\{$~\WRITE$(r_{j_1}, v_{j_1}, \langle date_{j_1}, j_1\rangle),~\ldots,$
     \WRITE$(r_{j_x}, v_{j_x}, \langle date_{j_x}, j_x\rangle),$ \\

\>\>$~~~~~~~~~~~~~~~~~~~~~~~~~~~~~$
   \SYNC$(j_{x+1}),~\ldots,$ \SYNC$(j_{y})~\}$ {\bf is scd-delivered do}\\

\line{Snap-from-SC-05} 
\>\> {\bf for} \= {\bf  each} 
   $r$ such that  \WRITE$(r,-,-)$ $\in$  scd-delivered message set {\bf do}\\

\line{Snap-from-SC-06} 
\>\>\> {\bf let} $\langle date, writer\rangle$ be the greatest timestamp
in the messages  \WRITE$(r,-,-)$;\\

\line{Snap-from-SC-07} 
\>\>\> {\bf if} \=  $(\tsa_i[r] \lexts \langle date,writer \rangle $) \\

\line{Snap-from-SC-08}
\>\>\>\>  {\bf then} \=

   {\bf let} $v$ the value in  \WRITE$(r,-,\langle date,writer \rangle)$;\\

\line{Snap-from-SC-09}
 \>\>\>\>\>
 $reg_i[r]\leftarrow v$; $\tsa_i[r]\leftarrow  \langle date,writer \rangle$\\

\line{Snap-from-SC-10}
\>\>\> {\bf end if}\\

\line{Snap-from-SC-11}
\>\> {\bf end for};\\

\line{Snap-from-SC-12}  \>\>
              {\bf if} $\exists \ell~:~j_\ell=i$ 
                {\bf then} $\done_i \leftarrow \ttrue$ {\bf end if}. 

\end{tabbing}
\end{minipage}
}
  \caption{Construction of an MWMR  snapshot object 
           $\CAMP_{n,t}[\mbox{\SCD-broadcast}]$ (code for $p_i$)}
\label{algo:snapshot-from-SCD}
}
\end{algorithm}

\paragraph{Algorithm~\ref{algo:snapshot-from-SCD}: write operation}
(Lines~\ref{Snap-from-SC-03}-\ref{Snap-from-SC-04}) When a process
$p_i$ wants to assign a value $v$ to $\REG[r]$, it invokes
$\REG.\wwrite(r,v)$.  This operation is made up of two instances of
the communication pattern.  The first one is a re-synchronization
(line~\ref{Snap-from-SC-03}), as in the snapshot operation, whose side
effect is here to provide $p_i$ with an up-to-date value of
$\tsa_i[r].date$.
In the second instance of the communication pattern, 
$p_i$ associates the timestamp $\langle \tsa_i[r].date+1,i\rangle$ with $v$,
and scd-broadcasts the data/control
message \WRITE$(r, v, \langle \tsa_i[r].date+1,i\rangle)$.
In addition to informing the other
processes on its write of $\REG[r]$, this message \WRITE$()$ acts as a
re-synchronization message, exactly as a message \SYNC$(i)$.  When
this synchronization terminates (i.e., when the Boolean $\done_i$ is
set to ${\ttrue}$), $p_i$ returns from the write operation.

\paragraph{Algorithm~\ref{algo:snapshot-from-SCD}:
  scd-delivery of a set of messages}
When $p_i$ scd-delivers a message set, namely, \\
\centerline{
$\{$~\WRITE$(r_{j_1}, v_{j_1}, \langle date_{j_1}, j_1\rangle),~\ldots,$
     \WRITE$(r_{j_x}, v_{j_x}, \langle date_{j_x}, j_x\rangle),$ 
      \SYNC$(j_{x+1}),~\ldots,$ \SYNC$(j_{y})~\}$}\\ 
it first looks if there are messages \WRITE$()$.  If it is the case,
for each register $\REG[r]$ for which there are messages
\WRITE$(r,-,-)$ (line~\ref{Snap-from-SC-05}), $p_i$ computes the
maximal timestamp carried by these messages
(line~\ref{Snap-from-SC-06}), and updates accordingly its local
representation of $\REG[r]$
(lines~\ref{Snap-from-SC-07}-\ref{Snap-from-SC-10}).  Finally, if
$p_i$ is the sender of one of these messages (\WRITE$()$ or
\SYNC$()$), $done_i$ is set to $\ttrue$, which terminates $p_i$'s
re-synchronization (line~\ref{Snap-from-SC-12}).

\paragraph{Time and Message costs}
An invocation of  $\snapshot()$ involves one
invocation of $\scdbroadcast()$, while an  invocation of
$\wwrite()$ involves two such invocations. 
As $\scdbroadcast()$ costs $O(n^2)$ protocol
messages and $2\Delta$ time units,  $\snapshot()$
cost the same, and $\wwrite()$ costs the double.

\subsection{Proof of Algorithm~\ref{algo:snapshot-from-SCD}}
\label{sec:proof-snapshot}
As they are implicitly used in the  proofs that follow, let us recall
the properties of the \SCD-broadcast abstraction. The non-faulty processes
scd-deliver the same messages (exactly one each), and
each of them was scd-broadcast. As a faulty process behaves correctly
until it crashes, it scd-delivers a subset of the messages 
scd-delivered by the non-faulty processes. 

Without loss of generality, we assume that there is an initial write
operation issued by a non-faulty process.
Moreover, if  a process crashes in a snapshot operation, its snapshot is not
considered;  if  a process crashes in a write operation, its
write is considered only if the message \WRITE() it sent at
line~\ref{Snap-from-SC-04} is scd-delivered to at least one non-faulty process
(and by the Termination-2 property, at least to all non-faulty processes). 
Let us notice that  a message  \SYNC() scd-broadcast by a process $p_i$
does not modify the local variables of  the other processes.

\begin{lemma}
\label{lemma:snapshot-liveness}
If a non-faulty process invokes an operation, it returns from its invocation. 
\end{lemma}

\begin{proofL}
  Let $p_i$ be a non-faulty process that invokes a read or write  operation.
  By the Termination-1 property of \SCD-broadcast, it 
  eventually receives a message set containing the message
  \SYNC() or \WRITE() it sends at 
  line \ref{Snap-from-SC-02}, \ref{Snap-from-SC-03} or \ref{Snap-from-SC-04}.
  As all the statements associated with the scd-delivery of a message set
  (lines~\ref{Snap-from-SC-05}-\ref{Snap-from-SC-12}) terminate, 
  it follows that  the synchronization Boolean $\done_i$   is eventually
  set to $\ttrue$. Consequently,  $p_i$ returns from the invocation of its
  operation.
\renewcommand{\toto}{lemma:snapshot-liveness}
\end{proofL}

\paragraph{Extension of the relation $\lexts$}
The relation $\lexts$ is extended to  
a partial order on arrays of timestamps, denoted $ \le_{\tsa}$,
defined as follows: 
$tsa1[1..m] \le_{\tsa} tsa2[1..m] \stackrel{\mathit{def}}{=}
\forall r: (\tsa1[r] = \tsa2[r] \lor \tsa1[r] \lexts\ tsa2[r])$.
Moreover,  $\tsa1[1..m] <_{\tsa} \tsa2[1..m] \stackrel{\mathit{def}}{=}
(\tsa1[1..m] \le_{tsa} \tsa2[1..m])\wedge (\tsa1[1..m] \neq tsa2[1..m])$.

\paragraph{Definition}
Let $\TSA_i$ be the set of the array values  taken by $ts_i[1..m]$ at
line~\ref{Snap-from-SC-12} (end of the processing of  a message set
by process $p_i$). Let $\TSA=\cup_{1\leq i\leq n} \TSA_i$. 

\begin{lemma}
\label{lemma:clock-ordering}
The order $\le_{\tsa}$ is total on $\TSA$.
\end{lemma}

\begin{proofL}
  Let us first observe that, for any $i$,  all values in $\TSA_i$
  are totally ordered (this comes from $ts_i[1..m]$ whose entries can
  only increase,  lines~\ref{Snap-from-SC-07} and~\ref{Snap-from-SC-10}).  
  Hence, let $\tsa1[1..m]$ be an array value
  of $\TSA_i$, and  $\tsa2[1..m]$ an array value of $\TSA_j$, where $i\neq j$.

  Let us assume, by contradiction, that $\lnot (\tsa1 \le_{\tsa}
  \tsa2)$ and $\lnot (\tsa2 \le_{\tsa} \tsa1)$.  As $\lnot (\tsa1
  \le_{\tsa} \tsa2)$, there is a registers $r$ such that $\tsa2[r] <\tsa1[r]$.
  According to lines~\ref{Snap-from-SC-07}
  and~\ref{Snap-from-SC-09}, there is a message \WRITE$(r,-,  \tsa1[r])$
  received by $p_i$ when $\tsa_i = \tsa1$ and not received  
  by $p_j$ when $\tsa_j = \tsa2$ (because $\tsa2[r] <\tsa1[r]$).
  Similarly, there is a message
  \WRITE$(r',-, \tsa2[r'])$ received by $p_j$ when $\tsa_j = \tsa2$
  and not received by $p_i$ when $\tsa_i = \tsa1$.
  This situation contradicts the MS-Ordering property, from which we conclude
  that either $\tsa1 \le_{\tsa} \tsa2$ or $\tsa2 \le_{\tsa} \tsa1$. 
  \renewcommand{\toto}{lemma:clock-ordering} 
\end{proofL}

\paragraph{Definitions}
Let us associate a timestamp $ts(\wwrite(r, v))$ with each write
operation as follows. Let $p_i$ be the invoking process; $ts(\wwrite(r,v))$
is the timestamp of $v$ as defined by $p_i$ at
line~\ref{Snap-from-SC-04}, i.e., $\langle \tsa_i[r].date+1, i\rangle$.

Let $\op1$ and $\op2$ be any two operations. The relation
$\prec$ on the whole set of operations is defined as follows:
 $\op1 \prec \op2$ if $\op1$ terminated before $\op2$ started. 
It is easy to see that $\prec$ is a real-time-compliant partial order 
on all the operations.

\begin{lemma}
\label{lemma:snapshot-write-ordering}
No two distinct write operations on the same register $\wwrite1(r, v)$
and $\wwrite2(r, w)$ have the same timestamp, and $(\wwrite1(r, v)
\prec\wwrite2(r, w))$ $\Rightarrow$ $(ts(\wwrite1) \lexts ts(\wwrite2))$.
\end{lemma}

\begin{proofL}
Let $\langle date1,i\rangle$ and $\langle date2,j\rangle$ be the
timestamp of $\wwrite1(r, v)$ and $\wwrite2(r, w)$, respectively.  If
$i\neq j$, $\wwrite1(r, v)$ and $\wwrite2(r, w)$ have been produced by
different processes, and their timestamp differ at least in their
process identity.

So, let us consider that the operations have been issued by the same
process $p_i$, with $\wwrite1(r, v)$ first.  As $\wwrite1(r, v)$
precedes $\wwrite2(r, w)$, $p_i$ first invoked  $\scdbroadcast$
\WRITE$(r, v,\langle date1,i\rangle)$ (line~\ref{Snap-from-SC-04}) and
later \WRITE$(r, w,\langle date2,i\rangle)$. It follows that these
\SCD-broadcast invocations are separated by a local reset of the Boolean
$done_i$ at line~\ref{Snap-from-SC-04}.  Moreover, before the reset of
$\done_i$ due to the scd-delivery of
the message $\{ \ldots, $\WRITE$(r, v,\langle date1,i\rangle),\ldots\}$,
we have $\tsa_i[r].date_i\geq date1$
(lines~\ref{Snap-from-SC-06}-\ref{Snap-from-SC-10}).  Hence, we have
$\tsa_i[r].date\geq date1$ before the reset of $\done_i$
(line~\ref{Snap-from-SC-12}).  Then, due to the ``$+1$'' at
line~\ref{Snap-from-SC-04}, \WRITE $(r, w,\langle date2,i\rangle)$ is
such that $date2>date1$, which concludes the proof of the first part
of the lemma. 

Let us now consider that  $\wwrite1(r, v) \prec\wwrite2(r, w)$. If
$\wwrite1(r, v)$ and $\wwrite2(r, w)$ have been produced by the same process
we have $date1 < date2$ from the previous reasoning.
So let us assume that they have been produced by different processes
$p_i$ and $p_j$.
Before  terminating $\wwrite1(r, v)$ (when the Boolean $\done_i$ is set
$\ttrue$  at line~\ref{Snap-from-SC-12}), $p_i$  received a message set
$ms1_i$ containing the message \WRITE$(r, v,\langle date1,i\rangle)$. 
When $p_j$ executes  $\wwrite2(r, w)$, it first  invokes
$\scdbroadcast$ \SYNC$(j)$ at line~\ref{Snap-from-SC-03}. 
Because $\wwrite1(r, v)$ terminated before  $\wwrite2(r, w)$ started, 
this message \SYNC$(j)$ cannot belong to $ms1_i$.

Due to Integrity and Termination-2 of \SCD-broadcast, $p_j$ eventually
scd-delivers exactly one  message set $ms1_j$ containing
\WRITE$(r, v,\langle date1,i\rangle)$. Moreover, it also
scd-delivers  exactly one  message set $ms2_j$ containing
its own message \SYNC$(j)$.
On the the other side, $p_i$ scd-delivers exactly one  message
set $ms2_i$ containing the message \SYNC$(j)$.
It follows from the MS-Ordering property that, if  $ms2_j\neq ms1_j$, 
$p_j$ cannot scd-deliver $ms2_j$  before  $ms1_j$.
Then, whatever the case ($ms1_j=ms2_j$ or  $ms1_j$ is scd-delivered at
$p_j$ before $ms2_j$), it follows from the fact that the messages \WRITE$()$
are processed (lines~\ref{Snap-from-SC-05}-\ref{Snap-from-SC-11})
before the messages \SYNC$(j)$ (line~\ref{Snap-from-SC-12}), that we have
$\tsa_j[r]\geq \langle date1,i\rangle$ when $\done_j$ is set to $\ttrue$. 
It then follows from line~\ref{Snap-from-SC-04} that $date2>date1$,
which concludes the proof of the lemma. 
\renewcommand{\toto}{lemma:snapshot-write-ordering}
\end{proofL}

\paragraph{Associating timestamp arrays with operations}
Let us associate a timestamp array $\tsa(\op)[1..m]$ with each operation
$\op()$ as follows.
\begin{itemize}
\vspace{-0.2cm}
\item
  Case $\op()=\snapshot()$. Let $p_i$ be the invoking process;
  $\tsa(\op)$ is the value of $\tsa_i[1..m]$ when $p_i$ returns from the
  snapshot operation (line~\ref{Snap-from-SC-02}).  
\vspace{-0.2cm}
\item Case $\op()=\wwrite(r,v)$. 
  Let $\mmin_\tsa(\{A\})$, where $A$ is a set of array values,      
  denote the smallest array value of $A$ according to $<_\tsa$. 
  Let $\tsa(\op)\stackrel{\mathit{def}}{=}   
  \mmin_{\tsa}(\{\tsa[1..m] \in \TSA
                    \mbox{ such  that } ts(\op) \leq_{ts} \tsa[r]\})$.    
   Hence,  $\tsa(\op)$ is the first $\tsa[1..m]$ of $\TSA$, that reports
   the operation $\op()=\wwrite(r,v)$.        
\end{itemize}

\begin{lemma}
\label{lemma:snapshot-clock-ordering}
Let $\op$ and $\op'$ be two distinct operations such that $\op
\prec\op'$. We have $\tsa(\op) \le_{\tsa} \tsa(\op')$. Moreover, if $\op'$
is a write operation, we have $\tsa(\op) <_{\tsa} \tsa(\op')$.
\end{lemma}

\begin{proofL}
  Let $p_i$ and $p_j$ be the processes that performed $\op$ and 
  $\op'$, respectively.
  Let \SYNC$_j$ be the \SYNC$(j)$ message sent by $p_j$ (at
  line~\ref{Snap-from-SC-02} or~\ref{Snap-from-SC-03}) during the
  execution of $\op'$.  Let $\mathit{term\_\tsa_i}$ be the
  value of $\tsa_i[1..m]$ when $\op$ terminates
  (line~\ref{Snap-from-SC-02} or~\ref{Snap-from-SC-04}), and
  $\mathit{sync\_\tsa_j}$
  the value of $\tsa_j[1..m]$ when $\mathit{done_j}$ becomes true for the first
  time after $p_j$ sent \SYNC$_j$ (line~\ref{Snap-from-SC-01}
  or~\ref{Snap-from-SC-03}). Let us notice that $\mathit{term\_\tsa_i}$
  and  $\mathit{sync\_\tsa_j}$ are elements of the set $\TSA$.

  According to lines~\ref{Snap-from-SC-07} and~\ref{Snap-from-SC-10}, for
  all $r$, $\tsa_i[r]$ is the largest timestamp carried by
  a message \WRITE$(r,v,-)$
  received by $p_i$ in a message set before $\op$ terminates.  Let $m$ be a
  message such that there is a set $sm$ scd-delivered by $p_i$ before it
  terminated $\op$. As $p_j$ sent \SYNC$_j$ after $p_i$ terminated,
  $p_i$ did not receive any set containing \SYNC$_j$ before it
  terminated $\op$. By the properties Termination-2 and MS-Ordering, $p_j$
  received message $m$ in the same set as \SYNC$_j$ or in a message
  set $sm'$ received before the set containing \SYNC$_j$. Therefore,
  we have $\mathit{term\_\tsa_i} \le_{\tsa} \mathit{sync\_\tsa_j}$.
  
  If $\op$ is a snapshot operation, then $\tsa(\op) =
  \mathit{term\_\tsa_i}$. Otherwise, $\op()=\wwrite(r,v)$. As $p_i$ has
  to wait until it processes a set of messages including  its \WRITE() message
  (and executes line~\ref{Snap-from-SC-12}),
   we have $ts(\op) \lexts \mathit{term\_\tsa_i}[r]$.
   Finally, due to the fact that $\mathit{term\_\tsa_i}\in \TSA$ and
   Lemma~\ref{lemma:clock-ordering}, we have
   $\tsa(\op)\le_{\tsa}\mathit{term\_\tsa_i}$.

  If $\op'$ is a snapshot operation, then $\mathit{sync\_\tsa_j} = \tsa(\op')$
   (line~\ref{Snap-from-SC-02}).
  Otherwise, $\op()=\wwrite(r,v)$ and thanks to the $+1$ in
  line~\ref{Snap-from-SC-04},
  $\mathit{sync\_\tsa_j}[r]$ is strictly smaller than $\tsa(\op')[r]$
  which, due to Lemma~\ref{lemma:clock-ordering}, implies
  $\mathit{sync\_\tsa_j} <_{\tsa} \tsa(\op')$.

  It follows that, in all cases, we have
  $\tsa(\op) \le_{\tsa} \mathit{term\_\tsa_i} \le_{\tsa} \mathit{sync\_\tsa_j}
                                                \le_{\tsa} \tsa(\op')$
  and if $\op'$ is a write operation, we have $\tsa(\op) \le_{\tsa}
  \mathit{term\_\tsa_i} \le_{\tsa} \mathit{sync\_\tsa_j} <_{\tsa} \tsa(\op')$,
  which concludes the proof of the lemma.
  \renewcommand{\toto}{lemma:snapshot-clock-ordering}
\end{proofL}


The previous lemmas allow the operations to be linearized (i.e.,
totally ordered in an order compliant with both the sequential
specification of a register, and their real-time occurrence order)
according to a total order extension of the reflexive and transitive
closure of the $\rightarrow_{lin}$ relation defined thereafter.
\begin{definition}
  Let $\op, \op'$ be two operations. We define the $\rightarrow_{lin}$ relation
  by $\op \rightarrow_{lin} \op'$ if one of the following properties holds:
  \begin{itemize}
  \item\vspace{-0.2cm}
    $\op \prec \op'$,
  \item\vspace{-0.2cm}
    $tsa(\op) <_{tsa} tsa(\op')$,
  \item\vspace{-0.2cm} $tsa(\op) = tsa(\op')$, $op$ is a write operation
    and $\op'$ is a snapshot operation,
  \item\vspace{-0.2cm} $tsa(\op) = tsa(\op')$, $\op$ and $\op'$ are two
    write operations on the same register and $ts(\op) \lexts ts(\op')$,
  \end{itemize}
\end{definition}

\begin{lemma}
 \label{lemma:snapshot-safety}
 The snapshot object built by Algorithm~{\em{\ref{algo:snapshot-from-SCD}}}
 is linearizable.
\end{lemma}

\begin{proofL}
  We recall the definition of the $\rightarrow_{lin}$ relation: $\op
  \rightarrow_{lin} \op'$ if one of the following properties holds:
  \begin{itemize}
  \item\vspace{-0.2cm}
    $\op \prec \op'$,
  \item\vspace{-0.2cm}
    $tsa(\op) <_{tsa} tsa(\op')$,
  \item\vspace{-0.2cm} $tsa(\op) = tsa(\op')$, $op$ is a write operation
    and $\op'$ is a snapshot operation,
  \item\vspace{-0.2cm} $tsa(\op) = tsa(\op')$, $\op$ and $\op'$ are two
    write operations on the same register and $ts(\op) \lexts ts(\op')$,
  \end{itemize}
  We define the $\rightarrow_{lin}^\star$ relation as the reflexive
  and transitive closure of the $\rightarrow_{lin}$ relation.

  Let us prove that the $\rightarrow_{lin}^\star$ relation is a
  partial order on all operations. Transitivity and reflexivity are
  given by construction.  Let us prove antisymmetry. Suppose there are
  $\op_0, \op_2, ..., \op_m$ such that $\op_0 = \op_m$ and $\op_i
  \rightarrow_{lin} \op_{i+1}$ for all $i<m$.  By
  Lemma~\ref{lemma:snapshot-clock-ordering}, for all $i<m$, we have
  $tsa(\op_i) \le_{tsa} tsa(\op_{i+1})$, and $tsa(\op_m) =
  tsa(\op_{0})$, so the timestamp array of all operations are the
  same.  Moreover, if $\op_i$ is a snapshot operation, then $\op_i
  \prec \op_{(i+1) \% m}$ is the only possible case
  ($\%$ stands for ``modulo'') , and by
  Lemma~\ref{lemma:snapshot-clock-ordering} again, $\op_{(i+1) \% m}$
  is a snapshot operation. Therefore, only two cases are possible.
  
  \begin{itemize}
    \vspace{-0.2cm}
  \item Let us suppose that all the $\op_i$ are snapshot operations
    and for all $i$, $\op_i \prec \op_{(i+1) \% m}$. As $\prec$ is a
    partial order relation, it is antisymmetric, so all the $\op_i$
    are the same operation.
    \vspace{-0.2cm}
  \item Otherwise, all the $\op_i$ are write operations. By
    Lemma~\ref{lemma:snapshot-clock-ordering}, for all $\op_i
    \not\prec \op_{(i+1) \% m}$.  The operations $\op_i$ and
    $\op_{i+1\% m}$ are ordered by the fourth point, so they are write
    operations on the same register and $ts(\op_i) \lexts
    ts(\op_{i+1\% m})$. By antisymmetry of the $\lexts$ relation, all
    the $\op_i$ have the same timestamp, so by
    Lemma~\ref{lemma:snapshot-write-ordering}, they are the same
    operation, which proves antisymmetry.
  \end{itemize}
  Let $\le_{lin}$ be a total order extension of
  $\rightarrow_{lin}^\star$. Relation $\le_{lin}$ is real-time
  compliant because $\rightarrow_{lin}^\star$ contains $\prec$.
  
  Let us consider a snapshot operation $\op$ and a register $r$ such
  that $tsa(\op)[r] = \langle date1, i \rangle$. According to
  line~\ref{Snap-from-SC-04}, it is associated to the value $v$ that
  is returned by $\rread1()$ for $r$, and comes from a
  \WRITE$(r,v,\langle date1, i \rangle)$ message sent by a write
  operation $\op_r = \wwrite(r,v)$.  By definition of $tsa(\op_r)$, we have
  $tsa(\op_r) \le_{tsa} tsa(\op)$ (Lemma~\ref{lemma:snapshot-clock-ordering}), 
  and therefore $\op_r \le_{lin}
  \op$. Moreover, for any different write operation $\op'_r$ on $r$,
  by Lemma~\ref{lemma:snapshot-write-ordering}, $ts(\op'_r) \neq
  ts(\op_r)$. If $ts(\op'_r) \lexts ts(\op_r)$, then $\op'_r \le_{lin}
  \op_r$. Otherwise, $tsa(\op) <_{tsa} tsa(\op'_r)$, and (due to the first item
  of the definition of  $\rightarrow_{lin}$) we have  $\op \le_{lin}  \op'_r$.
  In both cases, the value written by $\op_r$ is the last
  value written on $r$ before $\op$, according to $\le_{lin}$.
  \renewcommand{\toto}{lemma:snapshot-safety}
\end{proofL}

\begin{theorem}
\label{theorem:proof-snapshot}
Algorithm~{\em{\ref{algo:snapshot-from-SCD}}} builds an {\em MWMR} atomic
snapshot object in the model $\CAMP_{n,t}[\mbox{\em \SCD-broadcast}]$. The
operation $\snapshot$ costs one  {\em SCD-broaddast},  the $\wwrite()$
operation costs two.
\end{theorem}

\begin{proofT}
  The proof follows from
  Lemmas~\ref{lemma:snapshot-liveness}-\ref{lemma:snapshot-safety}.
  The cost of the operation $\snapshot()$  follows from
  line~\ref{Snap-from-SC-01}, and the one of $\wwrite()$  follows from
  lines~\ref{Snap-from-SC-03}-\ref{Snap-from-SC-04}. 
\renewcommand{\toto}{theorem:proof-snapshot}
\end{proofT}


\paragraph{Comparison with other algorithms}
Interestingly, Algorithm~\ref{algo:snapshot-from-SCD} is more
efficient (from both time and message point of views) than the stacking
of a read/write snapshot algorithm running on top of a message-passing
emulation of a read/write atomic memory (such a stacking would costs
$O(n^2\log n)$
messages and $O(n\Delta)$ time units, see Section~\ref{sec:rw-snapshot}).

\paragraph{Sequentially consistent snapshot object}
When considering Algorithm~\ref{algo:snapshot-from-SCD}, let us suppress
line~\ref{Snap-from-SC-01} and line~\ref{Snap-from-SC-03}
(i.e.,  the messages \SYNC\ are suppressed). 
The resulting algorithm implements a sequentially consistent snapshot object. 
This results from the suppression of the real-time compliance due to the
messages \SYNC.
The operation $\snapshot()$ is purely local, hence its cost is $0$.
The cost of the operation $\wwrite()$ is one SCD-broadcast, i.e.,
$2\Delta$ time units and $n^2$ protocol messages. 
The proof of this algorithm is left to the reader.

\section{SCD-broadcast in Action (its Power): Counter Object} 
\label{sec:SCD-power-counter}

\paragraph{Definition}
Let a {\it counter} be an object which can be manipulated by three
parameterless operations: $\increase()$, $\decrease()$, and $\rread()$.
Let $C$ be a counter.  From a sequential specification point of view
$C.\increase()$ adds $1$ to $C$, $C.\decrease()$ subtracts $1$ from
$C$, $C.\rread()$ returns the value of $C$.  As indicated in the
Introduction, due to its commutative operations, this object is a good
representative of a class of CRDT objects ({\it conflict-free replicated
data type} as defined in~\cite{SPBZ11}).

\begin{algorithm}[h!]
\centering{\fbox{
\begin{minipage}[t]{150mm}
\footnotesize 
\renewcommand{\baselinestretch}{2.5}
\resetline
\begin{tabbing}
aaa\=aa\=aaa\=aaaaaa\=\kill

{\bf operation} $\increase()$ {\bf is}\\

\line{at-counter-01}  \>\> $\done_i \leftarrow \ffalse$;
    $\scdbroadcast$ \PLUS $(i)$;  $\wait (\done_i)$;\\
\line{at-counter-02}  \>\> $\return ()$. \\~\\

{\bf operation} $\decrease()$ {\bf is} the
same as  $\increase()$ where \PLUS$(i)$ is replaced by   \MINUS$(i)$.\\~\\

{\bf operation} $\rread()$ {\bf is}\\

\line{at-counter-03}  \>\> $\done_i \leftarrow \ffalse$;
$\scdbroadcast$ \SYNC $(i)$;  $\wait (\done_i)$;\\
\line{at-counter-04}  \>\> $\return(counter_i)$.\\~\\

{\bf when the message set} \= 
$\{$~\PLUS$(j_1),~\ldots,$\MINUS$(j_x),~\ldots,$
   \SYNC$(j_{y}),~\ldots~\}$ {\bf is scd-delivered do}\\

\line{at-counter-05} 
\>\> {\bf let} $p~~~=$ number of messages \PLUS() in the message set;\\

\line{at-counter-06} 
\>\> {\bf let} $m~=$ number of messages \MINUS() in the message set;\\ 

\line{at-counter-07} 
\>\> $counter_i \leftarrow counter_i + p - m$;\\

\line{at-counter-08}  \>\>
              {\bf if} $\exists \ell~:~j_\ell=i$ 
                {\bf then} $\done_i \leftarrow \ttrue$ {\bf end if}. 

\end{tabbing}
\end{minipage}
}
\caption{Construction of an atomic counter  in
           $\CAMP_{n,t}[\mbox{\SCD-broadcast}]$ (code for $p_i$)}
\label{algo:atomic-counter}
}
\end{algorithm}

\paragraph{An algorithm satisfying linearizability}
Algorithm~\ref{algo:atomic-counter} implements an atomic counter $C$.
Each process manages a local copy of it denoted $counter_i$.
The text of the algorithm is self-explanatory.

The operation $\rread()$ is similar to the operation $\snapshot()$ of
the snapshot object.  Differently from the $\wwrite()$ operation on a
snapshot object (which requires a synchronization message \SYNC() and
a data/synchronization message \WRITE$()$), the update operations
$\increase()$ and $\decrease()$ require only one data/synchronization
message \PLUS() or \MINUS().  This is the gain obtained from the fact
that, from a process $p_i$ point of view, the operations $\increase()$
and $\decrease()$ which appear between two consecutive of its
$\rread()$ invocations are commutative.

\begin{lemma}
\label{lemma:counter-liveness}
If a non-faulty process invokes an operation, it returns from its invocation. 
\end{lemma}

\begin{proofL}
  Let $p_i$ be a non-faulty process that invokes an $\increase()$,
  $\decrease()$ or $\rread()$ operation.  By the Termination-1
  property of \SCD-broadcast, it eventually receives a message set
  containing the message \PLUS(), \MINUS() or \SYNC() it sends at line
  \ref{at-counter-01} or \ref{at-counter-03}.  As all the statements
  associated with the scd-delivery of a message set
  (lines~\ref{Snap-from-SC-05}-\ref{Snap-from-SC-08}) terminate, it
  follows that the synchronization Boolean $\done_i$ is eventually set
  to $\ttrue$. Consequently, $p_i$ returns from the invocation of its
  operation.  \renewcommand{\toto}{lemma:counter-liveness}
\end{proofL}

\begin{definition}
  Let $\op_i$ be an operation performed by $p_i$. We define
  $\past(\op_i)$ as a set of messages by:
  \begin{itemize}
  \item\vspace{-0.2cm}
    If $\op_i$ is an $\increase()$ or $\decrease()$
    operation, and $m_i$ is the message sent during its execution at
    line~{\em\ref{at-counter-01}}, then $\past(\op_i) = \{m : m \mapsto m_i\}$.
  \item\vspace{-0.2cm}
    If $\op_i$ is a $\rread()$ operation, then
    $\past(\op_i)$ is the union of all sets of messages
    $\scddelivered$ by $p_i$ before it executed
    line~{\em\ref{at-counter-04}}.
  \end{itemize}
\noindent
  We define the $\rightarrow_{lin}$ relation by $\op \rightarrow_{lin}
  \op'$ if one of the following conditions hold:
  \begin{itemize}
  \item\vspace{-0.2cm} $\past(\op)\varsubsetneq \past(\op')$;
  \item\vspace{-0.2cm} $\past(\op) = \past(\op')$, $\op$ is an
    $\increase()$ or a $\decrease()$ operation and $\op'$ is a
    $\rread()$ operation.
  \end{itemize}
\end{definition}

\begin{lemma}
 \label{lemma:counter-safety}
 The counter object built by Algorithm~{\em{\ref{algo:atomic-counter}}}
 is linearizable.
\end{lemma}

\begin{proofL}
  Let us prove that $\rightarrow_{lin}$ is a strict partial order
  relation.  Let us suppose $\op \rightarrow_{lin} \op'
  \rightarrow_{lin} \op''$. If $\op'$ is a $\rread()$ operation, we
  have $\past(\op)\subseteq \past(\op')\varsubsetneq \past(\op'')$. If
  $\op'$ is an $\increase()$ or a $\decrease()$ operation, we have
  $\past(\op)\varsubsetneq \past(\op')\subseteq \past(\op'')$.  In both
  cases, we have $\past(\op)\varsubsetneq \past(\op'')$, which proves
  transitivity as well as antisymmetry and irreflexivity since it is
  impossible to have $\past(\op)\varsubsetneq \past(\op)$.

  Let us prove that $\rightarrow_{lin}$ is real-time compliant. Let
  $\op_i$ and $\op_j$ be two operations performed by processes $p_i$
  and $p_j$ respectively, and let $m_i$ and $m_j$ be the message sent
  during the execution of $\op_i$ and $\op_j$ respectively, on
  line~\ref{at-counter-01} or~\ref{at-counter-03}.  Suppose that
  $\op_i \prec \op_j$ ($\op_i$ terminated before $\op_j$ started).
  When $p_i$ returns from $\op_i$, by the waiting
  condition of line~\ref{at-counter-01} or~\ref{at-counter-03}, it has
  received $m_i$, but $p_j$ has not yet sent $m_j$.
  Therefore, $m_i \mapsto_i m_j$, and consequently  $m_j\notin
  \past(\op_i)$.  By the waiting condition during the execution of
  $\op_j$ (line~\ref{at-counter-01} or~\ref{at-counter-03}), we have
  $m_j \in \past(\op_j)$. By the Containment property of SCD-broadcast,
  we therefore have $\past(\op_i) \varsubsetneq \past(\op_j)$, so $\op_i
  \rightarrow_{lin} \op_j$.
  Let $\le_{lin}$ be a total order extension of
  $\rightarrow_{lin}^\star$. It is real-time
  compliant because $\rightarrow_{lin}^\star$ contains $\prec$.

  Let us now consider the value returned by a  $\rread()$ operation $\op$.
  Let $p$ be the number
  of \PLUS() messages in $\past(\op)$ and let $m$ be the number of
  \MINUS() messages in $\past(\op)$.  According to
  line~\ref{at-counter-01}, $\op$ returns the value of $counter_i$
  that is modified only at line~\ref{at-counter-07} and contains
  the value $p-m$, by commutativity of additions and substractions.
  Moreover, due to the definition of $\rightarrow_{lin}$, 
  all pairs composed of a $\rread()$ and an $\increase()$
  or $\decrease()$ operations are ordered by $\rightarrow_{lin}$,
  and consequently, 
  $\op$ has the same $\increase()$ and $\decrease()$ predecessors
  according to both $\rightarrow_{lin}$ and to $\le_{lin}$.  Therefore, the
  value returned by $\op$ is the number of times $\increase()$ has
  been called, minus the number of times $\increase()$ has been
  called, before $\op$ according to $\le_{lin}$, which concludes the
  lemma.
\renewcommand{\toto}{lemma:counter-safety}
\end{proofL}

\begin{theorem}
\label{theorem-atomic-counter}
Algorithm~{\em\ref{algo:atomic-counter}} implements an atomic counter. 
\end{theorem}

\begin{proofT}
  The proof follows from
  Lemmas~\ref{lemma:counter-liveness} and \ref{lemma:counter-safety}.
\renewcommand{\toto}{theorem-atomic-counter}
\end{proofT}

\paragraph{An algorithm satisfying sequential consistency}
The previous algorithm can be easily modified to obtain a
sequentially consistent counter. To this end,  a technique similar to
the one introduced in~\cite{AW94} can be used to allow 
the operations  $\increase()$ and $\decrease()$  to have a fast
implementation.
``Fast'' means here that these   operations are purely local: they do not
require the invoking process to wait in the algorithm implementing them.
Differently, the operation  $\rread()$ issued by a process $p_i$ 
cannot be fast, namely, all the previous
$\increase()$ and $\decrease()$ operations issued by $p_i$ must be applied
to its local copy of the counter for its invocation of $\rread()$  terminates
(this is the rule  known under the name ``read your writes'').

\begin{algorithm}[h!]
\centering{\fbox{
\begin{minipage}[t]{150mm}
\footnotesize 
\renewcommand{\baselinestretch}{2.5}
\resetline
\begin{tabbing}
aaa\=aa\=aaa\=aaaaaa\=\kill

{\bf operation} $\increase()$ {\bf is}\\

\line{sc-counter-01}  \>\> $lsc_i \leftarrow lsc_i+1$; \\
\line{sc-counter-02} \>\> $\scdbroadcast$ \PLUS $(i)$; \\
\line{sc-counter-03}  \>\> $\return ()$. \\~\\

{\bf operation} $\decrease()$ {\bf is} the
same as  $\increase()$ where \PLUS$(i)$ is replaced by   \MINUS$(i)$.\\~\\

{\bf operation} $\rread()$ {\bf is}\\
\line{sc-counter-04} \>\> $\wait(lsc_i=0)$;\\
\line{sc-counter-05}  \>\> $\return(counter_i)$.\\~\\

{\bf when the message set} \= 
$\{$~\PLUS$(j_1),~\ldots,$\MINUS$(j_x),~\ldots~\}$ {\bf is scd-delivered do}\\

\line{sc-counter-06} 
\>\> {\bf let} $p~~~=$ number of messages \PLUS() in the message set;\\

\line{sc-counter-07} 
\>\> {\bf let} $m~=$ number of messages \MINUS() in the message set;\\ 

\line{sc-counter-08} 
\>\> $counter_i \leftarrow counter_i + p - m$;\\

\line{sc-counter-09} 
\>\> {\bf let} $c~=$ number of messages \PLUS$(i)$ and
                      \MINUS$(i)$ in the message set;\\ 

\line{sc-counter-10}  \>\> $lsc_i \leftarrow lsc_i - c$.

\end{tabbing}
\end{minipage}
}
\caption{Construction of a seq. consistent counter  in
           $\CAMP_{n,t}[\mbox{\SCD-broadcast}]$ (code for $p_i$)}
\label{algo:aseq-consistent-counter}
}
\end{algorithm}

Algorithm~\ref{algo:aseq-consistent-counter} is the resulting algorithm.
In addition to $counter_i$, each process manages a local synchronization
counter $lsc_i$ initialized to $0$, 
which counts the number of  $\increase()$ and $\decrease()$
executed by $p_i$ and not locally applied to $counter_i$.
Only when $lsc_i$ is equal to $0$, $p_i$ is allowed to read $counter_i$.  

The cost of an operation $\increase()$ and $\decrease()$
is $0$ time units plus the $n^2$ protocol messages of the underlying
SCD-broadcast. The time cost of the operation $\rread()$ by a process
$p_i$ depends on the value of $lsc_i$.
It is $0$ when $p_i$ has  no ``pending'' counter operations.  

\paragraph{Remark}
As in~\cite{AW94}, using the same technique, it is possible to design
a sequentially consistent counter in which the operation   $\rread()$
is fast,  while the operations $\increase()$ and $\decrease()$ are not. 

\section{SCD-broadcast in Action (its Power): Lattice Agreement Task} 
\label{sec:SCD-power-LA}

\paragraph{Definition}

Let $S$ be a partially ordered set, and $\leq$ its partial order
relation.  Given $S'\subseteq S$, an upper bound of $S'$ is an element
$x$ of $S$ such that $\forall~ y\in S':~ y\leq x$.  The {\it least
  upper bound} of $S'$ is an upper bound $z$ of $S'$ such that, for
all upper bounds $y$ of $S'$, $z\leq y$.
$S$ is called a {\it semilattice} if all its finite subsets have a
least upper bound.  Let $\lub(S')$ denotes the least upper bound of $S'$.

Let us assume that each process $p_i$ has an input value $in_i$
that is an element of a semilattice $S$. 
The {\it lattice agreement} task was introduced in~\cite{AHR95}
and generalized in~\cite{FRRRV12}. 
It provides each process with an operation  denoted $\propose()$,
such that a process $p_i$ invokes $\propose(in_i)$
(we say that $p_i$ proposes $in_i$); this operation returns an element 
$z\in S$ (we say that it decides $z$). 
The task is defined by the following properties, where it is assumed that
each non-faulty process invokes  $\propose()$.
\begin{itemize}
\vspace{-0.2cm}
\item Validity.
If  process $p_i$ decides $out_i$,
we have $in_i \leq out_i \leq \lub(\{in_1,\ldots,in_n\})$.
\vspace{-0.2cm}
\item Containment. If $p_i$ decides $out_i$ and $p_j$ decides
  $out_j$, we have $out_i \leq out_j$ or $out_j \leq out_j$.  
\vspace{-0.2cm}
\item Termination. If a  non-faulty proposes a value, it decides a value.
\end{itemize}

\paragraph{Algorithm}
Algorithm~\ref{algo:lattice-agreement} implements the lattice agreement task. 
It is a very simple algorithm,  which uses one instance of the
communication pattern introduced in
Section~\ref{sec:communication-pattern}. 
The text of the algorithm is self-explanatory.

\begin{algorithm}[h!]
\centering{\fbox{
\begin{minipage}[t]{150mm}
\footnotesize 
\renewcommand{\baselinestretch}{2.5}
\resetline
\begin{tabbing}
aaa\=aa\=aaa\=aaaaaa\=\kill

{\bf operation} $\propose(in_i)$ {\bf is}\\
\line{LA-01}  \>\> $out_i \leftarrow in_i$;\\
\line{LA-02}  \>\>  $\done_i \leftarrow \ffalse$;
 $\scdbroadcast$ \MSG $(i,in_i)$; 
 $\wait (\done_i)$;\\
\line{LA-03}  \>\> $\return (out_i)$. \\~\\

{\bf when the message set} \= 
$\{$~\MSG$(j_1, v_{j_1}),~\ldots,$ \MSG$(j_x, v_{j_x})\}$
{\bf is scd-delivered do}\\

\line{LA-04} 
\>\> {\bf for} \= {\bf  each} 
  \MSG$(j,v)$ $\in$ the  scd-delivered message set {\bf do}
      $out_i \leftarrow out_i \cup v$  {\bf end for};\\

\line{LA-05}  \>\>
              {\bf if} $\exists \ell~:~j_\ell=i$ 
                {\bf then} $\done_i \leftarrow \ttrue$ {\bf end if}. 

\end{tabbing}
\end{minipage}
}
  \caption{Solving Lattice Agreement in 
           $\CAMP_{n,t}[\mbox{\SCD-broadcast}]$ (code for $p_i$)}
\label{algo:lattice-agreement}
}
\end{algorithm}

\begin{theorem}
\label{theorem-lattice}
Algorithm ~{\em\ref{algo:lattice-agreement}} solves the lattice agreement task. 
\end{theorem}

\begin{proofT}
The Termination property follows from the assumption that all non-faulty
processes  propose a value, lines~\ref{LA-02} and~\ref{LA-05}.
The Validity property follows directly from lines~\ref{LA-01} and~\ref{LA-04}.\\

As far as the Containment property is concerned we have the following.
Let us assume,  by contradiction, that there are two processes
$p_i$ ans $p_j$ such that we have neither $out_i\leq out_j$ nor 
$out_j\leq out_j$.This means that there is a
value $v\in out_i\setminus out_j$, and a value $v'\in out_j\setminus out_i$.
Let $ms_i$ and $ms'_i$ be the message sets (scd-delivered by $p_i$) which
contained $v$ and $v'$ respectively. As $v\in out_i$ and $v'\notin out_i$,
we have $ms_i \neq  ms'_i$, and $ms_i$ was scd-delivered before $ms'_i$.

Defining  similarly $ms_j$ (containing $v'$) and $ms'_j$
(containing $v$), we  have  $ms'_j \neq  ms_j$, and $ms'_j$ 
was scd-delivered before $ms_j$. It follows
(see Section~\ref{sec:properties-SCD}) that we have
$m \mapsto_i m'$ and $m'\mapsto_j m$, from which it follows that
$\mapsto~ = \cup_{1\leq x\leq n} \mapsto_x$ is not a partial order.
A contradiction with SCD-broadcast definition. 
\renewcommand{\toto}{theorem-lattice}
\end{proofT}

\paragraph{Remark 1}
SCD-broadcast can be built on top of read/write registers 
(see below Theorem \ref{theorem:sc-broadcast-from-RW}).
It follows that the combination of Algorithm~\ref{algo:lattice-agreement}
and Algorithm~\ref{algo:Sc-broadcast-from-snapshot} provides us with a pure
read/write algorithm solving the lattice agreement task.
As far as we known, this is the first algorithm solving 
lattice agreement, based only on  read/write registers.

\paragraph{Remark 2}
Similarly to the algorithms implementing  snapshot objects and counters
satisfying sequential consistency (instead of linearizability), 
Algorithm~\ref{algo:lattice-agreement} uses no message \SYNC().

Let us also notice the following.
Objects are specified by ``witness'' correct executions, which
are defined by sequential specifications. According to the
time notion associated  with these sequences
we have two consistency conditions:
linearizability (the same ``physical'' time
for all the objects)  or sequential consistency (a logical time is associated
with each object, independently from the other objects). 
Differently, as distributed tasks are defined by relations from input
vectors to output vectors (i.e., without  referring  to  specific execution
patterns or a time notion),  the notion of a consistency condition
(such as linearizability or sequential consistency) is meaningless for tasks.

\section{The Computability Power of SCD-broadcast  (its Limits)} 
\label{sec:SCD-computability}

This section presents an algorithm building the SCD-broadcast abstraction
on top of SWMR snapshot objects. (Such snapshot objects can be easily obtained
from MWMR snapshot objects.) 
Hence, it follows from (a) this algorithm,
(b) Algorithm~\ref{algo:SCD-broadcast-in-MP}, 
and  (c) the impossibility proof to build an atomic register  on top of
asynchronous message-passing systems  where $t\geq n/2$ process may crash
~\cite{ABD95},  that 
\SCD-broadcast cannot be implemented in
$\CAMP_{n,t}[t\geq n/2]$, and  snapshot objects
and \SCD-broadcast are computationally equivalent. 

\subsection{From snapshot to SCD-broadcast}

\paragraph{Shared objects}
The shared memory is composed of two SWMR snapshot objects.
Let $\epsilon$ denote the empty sequence. 
\begin{itemize}
\vspace{-0.2cm}
\item
$\SENT[1..n]$: is a snapshot object, initialized to
$[\emptyset, \ldots,\emptyset]$, such that 
$\SENT[i]$ contains the messages scd-broadcast by $p_i$. 
\vspace{-0.2cm}
\item
$\SETSEQ[1..n]$: is a snapshot object,  initialized to
  $[\epsilon, \ldots,\epsilon]$, such that $\SETSEQ[i]$ contains
  the sequence of the sets of messages scd-delivered by $p_i$. 
\end{itemize}
The notation $\oplus$ is used for the concatenation of a message set
at the end of a sequence of message sets. 

\paragraph{Local objects}
Each process $p_i$ manages the following local objects.
\begin{itemize}
\vspace{-0.2cm}
\item $\sent_i$ is a  local copy of the snapshot object $\SENT$.
\vspace{-0.2cm}
\item  $\setseq_i$ is a  local copy of the snapshot object $\SETSEQ$. 
\vspace{-0.2cm}
\item $\todeliveri$ is an auxiliary variable whose aim is to contain the
  next message set that $p_i$ has to scd-deliver. 
\end{itemize}
The function $\members(set\_seq)$
returns the set of all the messages contained in  $set\_seq$.

\paragraph{Description of Algorithm~\ref{algo:Sc-broadcast-from-snapshot}}
When a process $p_i$ invokes  $\scdbroadcast(m)$, it adds $m$ to
$\sent_i[i]$ and $\SENT[i]$ to inform all the processes 
on the scd-broadcast of $m$. It then invokes the internal procedure
$\progress()$ from which it exits once it has
a set containing $m$ (line~\ref{nRW-01}). 

A background task $T$ ensures that all messages will be
scd-delivered (line~\ref{nRW-02}). This task invokes repeatedly
the internal procedure $\progress()$.
As, locally, both the application process and the underlying task $T$ can
invoke  $\progress()$, which accesses the local variables of $p_i$,
those variables  are protected by a local fair mutual exclusion algorithm
providing the operations $\entermutex()$ and  $\exitmutex()$
(lines~\ref{nRW-03} and~\ref{nRW-11}).

\begin{algorithm}[h!]
\centering{\fbox{
\begin{minipage}[t]{150mm}
\footnotesize 
\renewcommand{\baselinestretch}{2.5}
\resetline
\begin{tabbing}
aaa\=aa\=aaa\=aaaaaa\=\kill

{\bf operation} $\scdbroadcast(m)$ {\bf is}\\

\line{nRW-01} \> $\sent_i[i]  \leftarrow \sent_i[i] \cup \{m\}$;
                 $\SENT.\wwrite(\sent_i[i])$;  $\progress()$.~\\~\\


\line{nRW-02} \> {\bf background task} $T$ {\bf is}
                 {\bf repeat forever} $\progress()$  {\bf end repeat}.\\~\\


{\bf procedure} $\progress()$  {\bf is}\\

\line{nRW-03} \>\> $\entermutex()$;\\

\line{nRW-04} \>\>  $\catchup()$;\\

\line{nRW-05} \>\>  $\sent_i \leftarrow \SENT.\snapshot()$;\\

\line{nRW-06} \>\>  $\todeliveri \leftarrow
  (\cup_{1\leq j \leq n}~\sent_i[j]) \setminus \members(\setseq_i[i])$;\\

\line{nRW-07} \>\> {\bf if} \= $(\todeliveri \neq\emptyset)$\\

\line{nRW-08} \>\>\>
     {\bf then} \= $\setseq_i[i] \leftarrow \setseq_i[i] \oplus \todeliveri$;
                   $\SETSEQ.\wwrite(\setseq_i[i])$;\\
     
\line{nRW-09} \>\>\>   $\scdeliver(\todeliveri)$ \\ 

\line{nRW-10} \>\>  {\bf end if};\\ 

\line{nRW-11} \> \> $\exitmutex()$.\\~\\

 {\bf procedure} $\catchup()$  {\bf is}\\

\line{nRW-12}  \>\> $\setseq_i  \leftarrow \SETSEQ.\snapshot()$;\\

\line{nRW-13}  \>\> {\bf while} \= $(\exists j, set:~ set
\mbox{ is the first set in } \setseq_i[j]:~ set \not\subseteq
\members(\setseq_i[i])$  {\bf do} \\

\line{nRW-14} \>\>\> $\todeliveri \leftarrow set \setminus
                      \members(\setseq_i[i])$;  \\%

\line{nRW-15}\>\>\>
       $\setseq_i[i] \leftarrow \setseq_i[i] \oplus \todeliveri$;
       $\SETSEQ.\wwrite(\setseq_i[i])$;\\

\line{nRW-16} \>\>\> $\scdeliver(\todeliveri)$  \\

\line{nRW-17}  \>\> {\bf end while}.

\end{tabbing}
\end{minipage}
}
\caption{An implementation of
  \SCD-broadcast on top of snapshot objects  (code for $p_i$)}
\label{algo:Sc-broadcast-from-snapshot}
}
\end{algorithm}

The procedure $\progress()$ first invokes the internal procedure
$\catchup()$, whose aim is to allow $p_i$ to scd-deliver sets of messages
which have been scd-broadcast and not yet locally scd-delivered. 

To this end, $\catchup()$ works as follows (lines~\ref{nRW-12}-\ref{nRW-17}).
Process $p_i$ first obtains a snapshot of $\SETSEQ$, and saves it in
$\setseq_i$  (line~\ref{nRW-12}).
This allows $p_i$ to know which message sets have been
scd-delivered by all the processes; $p_i$ then enters a ``while'' loop to
scd-deliver as many message sets as possible according to what was
scd-delivered by the other processes. 
For each process $p_j$ that has scd-delivered a message set $set$ containing
messages not yet scd-delivered by $p_i$  (predicate of line~\ref{nRW-13}), 
$p_i$ builds a set $\todeliver_i$ containing the messages in $set$ that it
has not  yet scd-delivered  (line~\ref{nRW-14}),  and locally scd-delivers it
(line~\ref{nRW-16}). This local scd-delivery needs to update accordingly
both $\setseq_i[i]$ (local update) and $\SETSEQ[i]$ (global update). 

When it returns from $\catchup()$, $p_i$ strives to scd-deliver messages
not yet scd-delivered by the other processes. To this end, it first obtains
a snapshot of $\SENT$, which it stores in $\sent_i$  (line~\ref{nRW-05}).
If there are messages that can be scd-delivered (computation of $\todeliver_i$
at line~\ref{nRW-06}, and predicate at line~\ref{nRW-07}), $p_i$
scd-delivers them and updates  $\setseq_i[i]$ and  $\SETSEQ[i]$
(lines~\ref{nRW-07}-\ref{nRW-09}) accordingly.

\subsection{Proof of Algorithm~\ref{algo:Sc-broadcast-from-snapshot}}

\begin{lemma}
\label{lemma-broadcast-validity-2}
If a process scd-delivers a set containing a message $m$, some process invoked
$\scdbroadcast(m)$. 
\end{lemma}

\begin{proofL}
  The proof follows directly from the text of the algorithm, which copies
  messages from $\SENT$ to $\SETSEQ$, without creating new messages.
\renewcommand{\toto}{lemma-broadcast-validity-2}
\end{proofL}

\begin{lemma}
\label{lemma-broadcast-integrity-2}
No process scd-delivers  the same message twice. 
\end{lemma}

\begin{proofL}
  Let us first observe that, due to lines \ref{nRW-07} and~\ref{nRW-15},
  all messages that are scd-delivered at a process $p_i$ have been  added to
  $\setseq_i[i]$. The proof then follows directly from (a) this observation, 
  (b) the fact that (due to the local mutual exclusion at each process)
  $\setseq_i[i]$ is updated consistently,
  and (c) lines \ref{nRW-06} and~\ref{nRW-14},
  which state that a message already scd-delivered (i.e., a message
  belonging to  $\setseq_i[i]$) cannot be added to $\todeliver_i$. 
\renewcommand{\toto}{lemma-broadcast-integrity-2}
\end{proofL}

\begin{lemma}
\label{lemma-broadcast-termination1-a2}
Any invocation of  $\scdbroadcast()$ by a non-faulty process $p_i$ terminates.
\end{lemma}

\begin{proofL}
The proof consists in showing that the internal procedure
$\progress()$ terminates. 
As the  mutex algorithm is assumed to be fair, process $p_i$ cannot block
forever at line~\ref{nRW-03}. Hence, $p_i$ invokes the 
internal  procedure $\catchup()$. It then 
issues first a snapshot invocation on $\SETSEQ$ and stores the 
value it  obtains  the value of $\setseq_i$.
There is consequently a finite number of message sets in $\setseq_i$.
Hence,  the ``while'' of lines~\ref{nRW-13}-\ref{nRW-17}
can be executed only a finite number of times, and it follows that 
any invocation of $\catchup()$ by a non-faulty process terminates. 
The same reasoning (replacing $\SETSEQ$ by $\SENT$) shows that
process $p_i$ cannot block forever when it executes the 
lines~\ref{nRW-05}-\ref{nRW-10} of the procedure $\progress()$. 
\renewcommand{\toto}{lemma-broadcast-termination1-a2}
\end{proofL}

\begin{lemma}
\label{lemma-broadcast-termination1-b2}
If a non-faulty process scd-broadcasts a message $m$,
it scd-delivers a message set containing $m$.
\end{lemma}

\begin{proofL}
Let $p_i$ be a non-faulty process that  scd-broadcasts a message $m$.
As it is non-faulty, $p_i$ adds $m$ to  $\SENT[i]$
and then invokes  $\progress()$  (line~\ref{nRW-01}).
As $m\in \SENT$, it is eventually added to $\todeliveri$
if not yet scd-delivered (line~\ref{nRW-06}),
and scd-delivered at line~\ref{nRW-09}, which concludes the proof
of the lemma. 
\renewcommand{\toto}{lemma-broadcast-termination1-b2}
\end{proofL}

\begin{lemma}
\label{lemma-broadcast-termination2-2}
If a non-faulty process scd-delivers a message $m$, every  non-faulty process
scd-delivers a message set containing $m$.
\end{lemma}

\begin{proofL}
Let us assume that a process scd-delivers a message set containing
a message $m$. It follows that the  process that invoked $\scdbroadcast(m)$ 
added $m$ to $\SENT$ (otherwise no process could scd-deliver $m$). 
Let $p_i$ be a correct process. 
It invokes $\progress()$ infinitely often  (line~\ref{nRW-02}).
Hence, there is a first execution of $\progress()$
such that $sent_i$ contains $m$   (line~\ref{nRW-05}).
If then follows from line~\ref{nRW-06}  that $m$ will be added to
$\todeliver_i$ (if not yet scd-delivered). If follows that $p_i$ will
scd-deliver a set of messages containing $m$ at line~\ref{nRW-09}.    
\renewcommand{\toto}{lemma-broadcast-termination2-2}
\end{proofL}


\begin{lemma}
\label{lemma-broadcast-sc-ordering-2}
Let $p_i$ be a process that scd-delivers a set $ms_i$ containing a
message $m$ and later scd-delivers a set $ms'_i$ containing a message
$m'$.  No process $p_j$ scd-delivers first a set $ms'_j$ containing $m'$
and later a set $ms_j$ containing $m$.
\end{lemma}

\begin{proofL}
  Let us consider two messages $m$ and $m'$. Due to
   total order property on the  operations on  the  snapshot object
  $\SENT$, it is possible to order the write operations
  of $m$ and $m'$ into $\SENT$. Without loss of generality, let us assume
  that $m$ is added to $\SENT$ before $m'$.
  We show that no process scd-delivers $m'$ before $m$.\footnote{Let us
    notice that it is possible that a process scd-delivers them in two
    different message sets, while another process scd-delivers them in
    the same set  (which does not contradicts the lemma).}

Let us consider a process $p_i$ that scd-delivers the message $m'$.
There are two cases. 
\begin{itemize}
\vspace{-0.2cm}
\item $p_i$ scd-delivers the message $m'$  at line~\ref{nRW-09}. 
Hence, $p_i$ obtained $m'$ from the snapshot object $\SENT$
(lines~\ref{nRW-05}-\ref{nRW-06}).
As $m$ was written in $\SENT$ before $m'$, we conclude that
$\SENT$ contains $m$. It then follows from line~\ref{nRW-06} that,
if $p_i$ has not scd-delivered $m$ before (i.e., $m$ is not in  
$\setseq_i[i]$), then $p_i$ scd-delivers it in the same set as $m'$. 
\vspace{-0.2cm}
\item $p_i$ scd-delivers the message $m'$  at line~\ref{nRW-16}.
  Due to the predicate used at line~\ref{nRW-13} to build a set
  of message to scd-deliver, this means that there is a process $p_j$
  that has previously scd-delivered a set of messages containing $m'$.\\
  Moreover, let us observe that the first time the  message $m'$ is copied
  from $\SENT$ to some $\SETSEQ [x]$ occurs at line~\ref{nRW-08}.
  As $m$ was written in $\SENT$ before $m'$, the corresponding process
  $p_x$ cannot see $m'$ and not $m$. 
  It follows from the previous item that $p_x$ has scd-delivered $m$
  in the same  message set (as the one including $m'$), or in a previous 
  message set. It then follows from the predicate of line~\ref{nRW-13} that
  $p_i$ cannot scd-delivers $m'$ before $m$.

  To summarize, the scd-deliveries of message sets in the procedure
  $\catchup()$  cannot violate the MS-Ordering property, which is established
  at lines~\ref{nRW-06}-\ref{nRW-10}. 
\end{itemize}
\vspace{-0.5cm} 
\renewcommand{\toto}{lemma-broadcast-sc-ordering-2}
\end{proofL}

\begin{theorem}
\label{theorem:sc-broadcast-from-RW}
Algorithm~{\em{\ref{algo:Sc-broadcast-from-snapshot}}}
implements the {\em \SCD-Broadcast} abstraction in the system model
$\CARW_{n,t}[t<n]$.
\end{theorem}

\begin{proofT}
  The proof follows from
  Lemma~\ref{lemma-broadcast-validity-2} (Validity), 
  Lemma~\ref{lemma-broadcast-integrity-2} (Integrity),
  Lemmas~\ref{lemma-broadcast-termination1-a2}
  and~\ref{lemma-broadcast-termination1-b2} (Termination-1), 
  Lemma~\ref{lemma-broadcast-termination2-2} (Termination-2), and
  Lemma~\ref{lemma-broadcast-sc-ordering-2} (MS-Ordering).
\renewcommand{\toto}{theorem:sc-broadcast-from-RW}
\end{proofT}

\section{Conclusion} 
\label{sec:conclusion}

\paragraph{What was the paper on?}
This paper has introduced a new communication abstraction, suited to
asynchronous message-passing systems where computing entities
(processes) may crash.  Denoted SCD-broadcast, it
allows processes to broadcast messages and deliver {\it sets of
  messages} (instead of delivering each message one after the other).
More precisely, if a process $p_i$ delivers a set of messages
containing a message $m$, and later delivers a set of messages
containing a message $m'$, no process $p_j$ can deliver a set of
messages containing $m'$ before a set of messages containing $m$.
Moreover, there is no local constraint imposed on the processing order
of the messages belonging to a same message set.  SCD-broadcast has
the following noteworthy features:
\begin{itemize}
\vspace{-0.2cm}
\item It  can be implemented in asynchronous message passing systems where
  any minority of processes may crash. Its costs are upper bounded by 
  twice the network latency (from a time point of view) and
  $(O(n^2)$ (from a message point of view). 
\vspace{-0.2cm}
\item Its computability power is the same as the one of atomic
  read/write register (anything that can be implemented in
  asynchronous read/write systems can be implemented with SCD-broadcast). 
\vspace{-0.2cm}
\item It promotes a communication pattern which is simple to use,
  when one has to implement concurrent objects defined by a sequential
  specification or distributed tasks. 
\vspace{-0.2cm}
\item When interested in the implementation of a concurrent object $O$,
  a simple weakening of the  SCD-broadcast-based atomic implementation 
  of $O$ provides us with an  SCD-broadcast-based  implementation 
  satisfying sequential consistency (moreover, the  sequentially consistent
   implementation is more efficient than the atomic one). 
\end{itemize}

\paragraph{On programming languages for distributed computing}
Differently from sequential computing for which there are plenty of
high level languages (each with its idiosyncrasies), there is no specific
language for distributed computing. Instead, addressing distributed
settings is done by the enrichment of sequential computing languages with
high level communication abstractions. When considering asynchronous systems
with process crash failures, {\it total order  broadcast} is one of them.
SCD-broadcast is a candidate to be one of them, when one has to
implement read/write solvable objects and distributed tasks.


\section*{Acknowledgments}
This work has been partially supported by the Franco-German DFG-ANR
Project 14-CE35-0010-02 DISCMAT (devoted to connections between
mathematics and distributed computing) and the French ANR project
16-CE40-0023-03 DESCARTES (devoted to layered and modular structures
in distributed computing).


\end{document}